\documentclass[epj]{svjour}

\usepackage{epsfig,amssymb,amsfonts,amsmath,mathtools,bm,color,graphicx,orcidlink,bm,braket,multirow,dsfont,mathtools,xcolor,slashed}

\RequirePackage[numbers,sort&compress]{natbib}
\journalname{Eur. Phys. J. C}

\renewcommand{\vec}[1]{\mbox{\boldmath$#1$\unboldmath}}
\newcommand{\Tch}{T_{\rm ch}}
\newcommand{\mq}{m_q}

\newcommand{\vp}{\varphi}

\newcommand{\be}{\begin{equation}}
\newcommand{\ee}{\end{equation}}
\newcommand{\bea}{\begin{eqnarray}}
\newcommand{\eea}{\end{eqnarray}}
\newcommand{\beas}{\begin{eqnarray*}}
\newcommand{\eeas}{\end{eqnarray*}}
\newcommand{\ds}{\displaystyle}
\newcommand{\vep}{{\bm p}}
\newcommand{\vek}{{\bm k}}
\newcommand{\veq}{{\bm q}}

\newcommand{\vex}{{\bm x}}
\newcommand{\vey}{{\bm y}}

\newcommand{\veA}{{\bm A}}

\newcommand{\muIR}{\mu_{\scalebox{0.6}{{\rm IR}}}}

\def\vec#1{{\bm #1}}

\newcommand{\vph}{\varphi}

\newcommand{\matrwf}{\phi}

\graphicspath{{Figs.dir/}}

\title{Confined but chirally and chiral spin symmetric hot matter}

\author{L. Ya. Glozman\inst{1}\thanks{E-mail: leonid.glozman@uni-graz.at}
 \and
 A. V. Nefediev\inst{2,3}\thanks{E-mail: a.nefediev@uni-bonn.de}
 \and
 R. F. Wagenbrunn\inst{1}\thanks{E-mail: robert@wagenbrunn.net}
}

\institute{Institute of Physics, University of Graz, A-8010 Graz, Austria
\and
Helmholtz-Institut f\"ur Strahlen- und Kernphysik, Universit\"at Bonn, D-53115 Bonn, Germany
\and
Center of Physics and Engineering of Advanced Materials, Instituto Superior T{\'e}cnico, 1049-001 Lisbon, Portugal
}


\date{}

\abstract{
We investigate properties of the quark--antiquark mesons at zero and finite temperature in the framework of a solvable chirally symmetric quark model. The interquark linearly rising interaction  is reminiscent of that derived in Coulomb gauge QCD, with the string tension being the only model parameter. We demonstrate that while the confining interaction induces spontaneous breaking of chiral symmetry at $T=0$, it gets restored at a temperature $\Tch\simeq 90$ MeV for the string tension fixed to provide the phenomenological value of the quark condensate. This temperature is similar to $\Tch\simeq 130$ MeV observed on the lattice in the chiral limit for $N_c=3$. The physical mechanism responsible for chiral symmetry restoration in the confining regime is Pauli blocking of the quark levels, required for the existence of a nonvanishing quark condensate, by thermal excitations of the quarks and antiquarks. Thus, above the chiral restoration temperature, meson-like states are chirally symmetric and approximately chiral spin symmetric. A crucial property of the confined meson-like light-light states above $\Tch$ is their size that exceeds drastically that in the chirally broken phase below $\Tch$. Heavy-heavy mesons nearly preserve their size irrespective of the temperature. Furthermore, the root-mean-square radius of the states with $J=0,1$ diverges in the chiral limit. This unexpected property must be a key to understanding unusual features of the hot QCD matter as observed at RHIC and LHC. Consequently, the confining but chirally symmetric matter above $\Tch$ can be considered as a dense system of very large and strongly overlapping meson-like states (``strings'').
}

\begin{document}

\maketitle

\section{Introduction}

As seen experimentally, the properties of the hot QCD matter
above the chiral restoration temperature $\Tch \simeq 155$ MeV differ radically from those of both a dilute hadron gas at low temperatures and a weakly interacting quark-gluon plasma (QGP) at very high temperatures. Hot QCD, as observed at BNL and CERN, is a highly collective and strongly interacting medium with
a small value of the ratio of the shear viscosity to the entropy density, $\eta/s$, and a small mean free path
of the effective constituents \cite{Heinz:2013th}. It has been established on the lattice that, when in equilibrium, this matter is not only chirally symmetric but also approximately chiral spin symmetric~\cite{Rohrhofer:2017grg,Rohrhofer:2019qwq,Rohrhofer:2019qal,Chiu:2023hnm}.
The chiral spin symmetry $SU(2)_{CS}$ --- a recently discovered new symmetry in QCD \cite{Glozman:2014mka,Glozman:2015qva} ---
is a symmetry of the colour charge and confining electric part of the QCD Lagrangian\footnote{This symmetry
was first observed on the lattice through a degeneracy of hadrons at $T=0$ upon an artificial truncation
of the near-zero modes of the Dirac operator \cite{Denissenya:2014poa,Denissenya:2014ywa}} (for a review see Ref.~\cite{Glozman:2022zpy}). It is
explicitly violated by magnetic interactions and the quark kinetic term. For an arbitrary number of quark flavours $N_f$, it can be extended to $SU(2N_F)$, that includes the chiral group
$SU(N_F)_R \times SU(N_F)_L \times U(1)_A$ as a subgroup. Observation of this
symmetry above $\Tch$ suggests that QCD is still in the confining
regime and the QCD partition function above $\Tch$ is not only chirally symmetric but also approximately $SU(2N_F)$ symmetric. The regime of QCD between the chiral restoration crossover and a very smooth crossover to QGP at, roughly, $3\Tch$ was tagged as ``stringy fluid''. There exist further lattice evidences, not related to symmetries, supporting
the existence of this intermediate regime of QCD above $\Tch$ \cite{Glozman:2022lda,Lowdon:2022xcl,Bala:2023iqu,Cohen:2024ffx}.
It still remains to be seen how the symmetries relevant for this intermediate regime and the properties
of the colour-singlet hadron-like states are related with the experimentally observed features of the hot matter such as collectivity, small ratio $\eta/s$, a very small mean free path of the effective constituents, and so on.

In this paper, we attempt to shed light on these puzzles employing a chiral quark model for QCD. Indeed, the situation should become more transparent in a combined large-$N_c$ and chiral limit. In this case, the quark loops are
suppressed and the $Z_{N_c}$ centre symmetry of the pure gauge action becomes exact in QCD with massless quarks. In particular, one unambiguously defines the deconfinement
temperature via the Polyakov loop \cite{Polyakov:1978vu} and then may address well-posed questions of the phases of the theory with confinement or deconfinement and spontaneously broken or restored chiral symmetry. A standard large-$N_c$ analysis \cite{Cohen:2023hbq} suggests that, while the above three regimes are connected by smooth crossovers in the world with $N_c=3$ and small but nonzero quark mass, they may become distinct
phases in the limit $N_c\to\infty$ and for a vanishing quark mass.

At present, it does not look feasible to get insight into this physics directly
from QCD. However, one may employ a solvable manifestly confining and chirally symmetric quark model in 3+1 dimensions suggested about 40 years ago \cite{Amer:1983qa,LeYaouanc:1983huv,LeYaouanc:1984ntu,Adler:1984ri,Kocic:1985uq,Bicudo:1989sh,Bicudo:1989si} that
has been actively studied in the literature till present days (see, for example, Refs.~\cite{Llanes-Estrada:1999nat,Bicudo:2002eu,Nefediev:2004by,Alkofer:2005ug,Wagenbrunn:2007ie}). Among many other properties, this model shares chiral spin symmetry of the confining interaction with QCD \cite{Glozman:2024xll}.
Within this model, the only interquark interaction mediated by the gluon field
is the linear "Coulomb-like" confining potential while all other possible
interactions, that are inherent in real QCD, are disregarded. Such simplifications are well
justified because the model still shares with real QCD its most crucial properties such as confinement as well chiral
symmetry and its spontaneous breaking in the vacuum.
At the same time, contrary to real QCD and its lattice implementation, the
model is tractable and consequently can provide a microscopic insight into
the crucial QCD phenomena.
This model is reminiscent of the celebrated 't~Hooft model for QCD in 1+1 dimensions in the large-$N_c$ limit \cite{tHooft:1974pnl,Kalashnikova:2001df}. Earlier, we derived and solved the finite-temperature mass-gap equation in this model and obtained a parameter-free relation between the chiral symmetry restoration temperature $\Tch$ and the chiral condensate at $T=0$ \cite{Glozman:2024xll},
\be
|\braket{\bar \psi \psi}|^{1/3} \simeq 2.75\;\Tch,
\ee
where the numerical factor on the right-hand side is determined by the physics of the linear confining ``Coulomb'' potential, which is the only interquark interaction within the model. Then, employing the phenomenological value of the quark condensate in QCD with $N_c=3$, $\braket{\bar\psi\psi}=-(250~\mbox{MeV})^3$, one obtains $\Tch \simeq 90$ MeV, in qualitative agreement with $\Tch \simeq 130$ MeV extracted on the lattice for QCD in the chiral limit and $N_c=3$ \cite{HotQCD:2019xnw}.\footnote{A similar result was obtained in Ref.~\cite{Quandt:2018bbu} in a different framework based on compactification of one spatial direction in Euclidean space.}. We have also clarified the physical mechanism of the chiral symmetry restoration: the thermal excitations of quarks and antiquarks
lead to Pauli blocking of the levels necessary for the formation of the quark condensate.

With a numerical solution of the mass-gap equation at $T=0$ at hand, the Bethe--Salpeter equation for quark-antiquark mesons in the vacuum can be formulated and solved --- there exists a vast literature on the subject. For example, in Ref.~\cite{Wagenbrunn:2007ie}, excited solutions of the Bethe--Salpeter equation were studied numerically,
and the obtained spectrum of mesons was demonstrated to exhibit effective restoration of chiral symmetry in the states with a large angular momentum. Additionally, emergence of the $SU(4) \times SU(4)$ symmetry of confinement (for $N_f=2$) was also observed.
In the meantime, in the employed model, the confining interaction remains operative at all temperatures. Then, while chiral symmetry is restored above $\Tch$, hadron-like states made of quarks are expected to persist and remain bound, so the system is still in the confining regime.\footnote{Notice that the 't~Hooft anomaly matching conditions \cite{tHooft:1979rat} and
the Coleman-Witten theorem \cite{Coleman:1980mx} are restricted to $T=0$ only when Lorentz invariance is manifest. They do not apply at $T>0$ and to dense hadronic systems. The same is true \cite{Glozman:2007tv,Glozman:2009sa} for the Casher argument \cite{Casher:1979vw}.}
Thus, in the present paper, we update the Bethe--Salpeter equation to accommodate for a finite
temperature, then solve it numerically for several $T$'s, and finally study the properties of the meson-like states above
$\Tch$. The most important results obtained are
\begin{itemize}
\item[(i)] the spectrum exhibits chiral symmetry and approximate $SU(4) \times SU(4)$ symmetry of confinement;
\item[(ii)] the light-light meson-like states acquire a much larger size at $T>\Tch$ than at $T<\Tch$; the strongest effect is observed for the states with the total spin $J=0,1$ that, in the strict chiral limit, become infinitely large even though they are still in the confining regime.
\end{itemize}
The latter property suggests that the stringy fluid matter is a dense system of long and strongly overlapping strings. This insight may facilitate further understanding of many experimental observations made for QCD above $\Tch$.

\section{Lessons from the Nambu--Jona-Lasinio model}

The Nambu--Jona-Lasinio (NJL) model described by the interaction Hamiltonian,
\be
H_{\rm int}^{\rm NJL}=-G\int d^3 x \left[(\bar \psi \psi)^2+(\bar \psi i\gamma_5\vec{\tau}\psi)^2\right],
\label{NJL}
\ee
was suggested long ago \cite{Nambu:1961tp,Nambu:1961fr} as a dynamical field-theory-inspired quark model for hadrons built in analogy with the theory of superconductivity \cite{Bardeen:1957kj,Bardeen:1957mv}; see also the dedicated review papers \cite{Vogl:1991qt,Klevansky:1992qe,Hatsuda:1994pi,Buballa:2003qv}. The model \eqref{NJL} is successful in describing the phenomenon of spontaneous breaking of chiral symmetry (SBCS) in the vacuum and is able to explain at a microscopic level the appearance of a gap in the spectrum of excitations provided by the dynamically generated quark mass --- here the analogy with the theory of superconductivity becomes particularly evident.
Indeed, in this model, the dynamical mass of the quark is related with the chiral condensate in the vacuum as
\be
M=-2G\braket{\bar \psi\psi}_0,
\label{gap1}
\ee
with the latter, in turn, evaluated as
\be
\braket{\bar\psi\psi}_0=-2N_c
\int\frac{d^3p}{(2\pi)^3} \frac{M}{\sqrt{p^2+M^2}},
\label{gap2}
\ee
where we work at $T=0$ and accordingly tag the chiral condensate with the corresponding subscript.
Equations \eqref{gap1} and \eqref{gap2} taken together result in a simple selfconsistency condition, known as the gap equation, for the effective quark mass $M$. The integral on the right-hand side of Eq.~\eqref{gap2} diverges quadratically and needs to be regularised, for example, by a sharp cut-off $\Lambda_\chi$. Then, for a fixed value of the cut-off $\Lambda_\chi$ and small $G$, the gap equation possesses only a trivial solution with $M=0$. However, if the strength of the effective interaction $G$ exceeds some critical level,
\be
G> \frac{\pi^2}{N_c\Lambda_\chi^2},
\ee
a solution with $M\neq 0$ emerges that describes a dynamically generated gap in the spectrum.

A finite temperature $T$ induces creation of particle-antiparticle excitations from the vacuum distributed in the relative momentum according to the Fermi-Dirac functions,
\begin{align}
n_p=\left(1 + e^{(\sqrt{p^2 + M^2}-\mu)/T}\right)^{-1}\mathop{\to}_{T\to\infty}\frac12,\nonumber\\[-2mm]
\label{FD}\\[-2mm]
\bar{n}_p=\left(1 + e^{(\sqrt{p^2 + M^2}+\mu)/T}\right)^{-1}\mathop{\to}_{T\to\infty}\frac12,\nonumber
\end{align}
with $n_p$ and $\bar{n}_p$ for the fermion and antifermion, respectively. Then the modified expression \eqref{gap2} for the chiral condensate at a finite $T$ reads
\be
\braket{\bar \psi\psi}=-\frac{N_c}{\pi^2}
\int_0^{\Lambda_\chi} {p}^2 dp\frac{M}{\sqrt{p^2 + M^2}}
(1-n_p-\bar{n}_p),
\label{gap4}
\ee
where the distribution functions in Eq.~\eqref{FD} effectively damp the interaction in the system, so that, for fixed values of $\Lambda_\chi$ and $G$, a nontrivial solution to the corresponding gap equation is lost at some critical temperature $\Tch$, even if such a solution existed at $T=0$. The physical mechanism behind chiral symmetry restoration is related to the fact that thermal excitations of quarks and antiquarks lead to Pauli blocking of the levels required for the formation of a nonvanishing quark condensate. For the properties of the chiral pion and $\sigma$-meson in the NJL model at finite temperature and baryon density see Refs.~\cite{Bernard:1987ir,Bernard:1987im}.

While the NJL model provides some valuable insight into spontaneous breaking of chiral symmetry in QCD it lacks confinement --- another aspect of QCD crucial for addressing the properties of excited hadrons that are in the spotlight of the present research. Thus we proceed to a chiral quark model with manifest confinement.

\section{Confining and chirally symmetric model in 3+1 dimensions}

\subsection{Symmetries of the QCD Hamiltonian in the Coulomb gauge}

The Hamiltonian of QCD in the Coulomb gauge in the chiral limit can be presented in the form \cite{Christ:1980ku}
\be
H_{\rm QCD}=H_E+H_B+\int d^3 x\;\psi^\dag(\vex)(-i{\bm\alpha}\cdot{\bm\nabla})\psi(\vex)+H_T+H_C,
\label{ham}
\ee
with
\be
H_T=-g\int d^3 x\;\psi^\dag(\vex){\bm\alpha}\cdot\veA^a(\vex)t^a \psi(\vex),
\ee
\be
H_C=\frac{g^2}2\int d^3\;x d^3 y\;J^{-1}\rho^a(\vex)F_{ab}(\vex,\vey) J\rho^b(\vey),
\label{coul}
\ee
where $J$ is the Faddeev-Popov determinant, $\rho^a(\vex)$ is the colour-charge density of quarks and gluons at the space point $\vex$, and $F_{ab}(\vex,\vey)$ is the ``Coulombic'' kernel. The quark kinetic and transverse (magnetic) part of the interaction of quarks with the gluonic field are chirally symmetric. Meanwhile, the confining ``Coulombic'' part is invariant under larger symmetry groups: $SU(2)_{CS}$, $SU(2N_F)$, and $SU(2N_F) \times SU(2N_F)$ \cite{Glozman:2022zpy}.

\subsection{Hamiltonian of the model}

The form of the instantaneous confining ``Coulombic'' term \eqref{coul} suggests that a sensible QCD-inspired confining and chirally symmetric model can be built in the form
\be
\begin{split}
H=&\int d^3x\;\psi^\dagger(\vex,t)\left(-i\vec{\alpha}\cdot
{\bm\nabla}+\beta \mq\right)\psi(\vex,t)\\
& + \frac{1}{2} \int d^3x\; d^3y\;\rho^a(\vex)K_{ab}(|\vex-\vey|)\rho^b(\vey),
\label{GNJL}
\end{split}
\ee
which includes the interaction of two quark colour charge densities,
$\rho^a=\psi^\dag\frac{\lambda^a}{2}\psi$, taken at the spatial points $\vex$ and $\vey$, via an instantaneous confining kernel,
\be
K_{ab}(|\vex-\vey|)=\delta_{ab}V_0(|\vex-\vey|).
\label{Kab}
\ee
This quark model has a rich history and has been widely discussed in the literature --- see, for example, Refs.~\cite{Amer:1983qa,LeYaouanc:1983huv,LeYaouanc:1984ntu,Adler:1984ri,Kocic:1985uq,Bicudo:1989sh,Bicudo:1989si,Bicudo:2002eu,Llanes-Estrada:1999nat,Nefediev:2004by,Alkofer:2005ug,Wagenbrunn:2007ie} and references therein.

A standard approach to solving the model implies a rainbow approximation for the dressed quark Green's function and a ladder approximation for the quark-antiquark Bethe--Salpeter equation. Such approximations are well justified in the large-$N_c$ limit and can be treated in the spirit of the 't~Hooft model for QCD in 1+1 dimensions \cite{tHooft:1974pnl}. A meaningful large-$N_c$ limit of the theory implies \emph{inter alia} that the strength of the confining potential scales with $N_c$ such that $C_F V_0(r)$, with $C_F=(N_C^2-1)/(2N_C)$ for the eigenvalue of the fundamental Casimir operator, reaches a finite limit as $N_c\to\infty$. Then, although in actual calculations the number of colours is set to three, the model still keeps the spirit of a large-$N_c$ theory. Therefore, the results obtained in this model can be directly confronted with those that follow from large-$N_c$ arguments --- see, for example, the discussion in Ref.~\cite{Cohen:2023hbq}.

We notice that qualitative conclusions deduced using the model \eqref{GNJL} are insensitive to a particular choice of the confining potential in Eq.~\eqref{Kab}. Meanwhile, in practical applications, a linearly rising confinement potential,
\be
V_{\rm conf}(r)=C_FV_0(r)=\sigma r,
\label{Vlin}
\ee
with $\sigma$ for the fundamental Coulomb string tension, is regarded as the most phenomenologically adequate choice. For example, in Refs.~\cite{Szczepaniak:2001rg,Feuchter:2004mk}, linear confinement was obtained as a result of variational calculations in quenched Coulomb-gauge QCD; see also Ref.~\cite{Nguyen:2024ikq} for recent insights and relevant references. Another attractive feature of the considered model with the linear potential \eqref{Vlin} is its aforementioned direct analogy with the 't~Hooft model for QCD in 1+1 dimensions in the large-$N_c$ limit \cite{tHooft:1974pnl} that was extensively studied in the axial (Coulomb) gauge, for example, in Refs.~\cite{Bars:1977ud,Li:1986gf,Kalashnikova:2001df,Glozman:2012ev}. In the two-dimensional model, the appearance of a linearly rising potential between quarks is a natural consequence of the form of the two-dimensional gluon propagator in the Coulomb gauge.

\subsection{Spontaneous breaking of chiral symmetry at $T=0$}
\label{sec:SBCS}

\begin{figure*}[t!]
\centering
\includegraphics[width=0.9\textwidth]{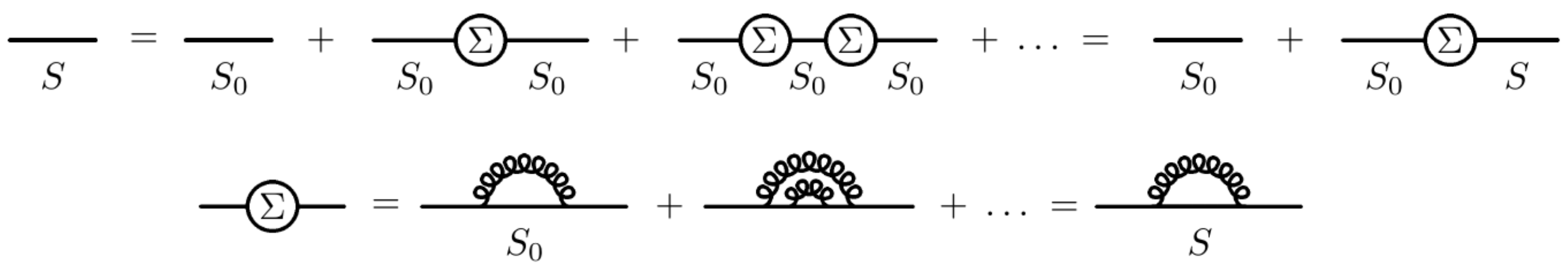}
\caption{Schematic representation of Eqs.~\eqref{Ds} and \eqref{Sigma0} derived in the rainbow approximation. The thin and thick solid lines are for the bare and dressed quark propagator, respectively, and the curly line is for the confining gluon propagator --- see Eq.~\eqref{FV} below for its explicit form that corresponds to linear confinement.}
\label{fig:dyson}
\end{figure*}

For a confining interquark interaction, the trivial chirally-symmetric vacuum is unstable~\cite{Amer:1983qa,LeYaouanc:1984ntu,Adler:1984ri,Bicudo:1989sh}. This instability can be studied with the help of the Bogoliubov-Valatin transformation of the quark field,
\be
\begin{split}
\psi(\vex,t)=\sum_{s=\uparrow,\downarrow}\int\frac{d^3p}{(2\pi)
^3}&e^{i\vep\vex}\Bigl[e^{-iE_pt}b_s(\vep)u_s(\vep)\\
&+e^{iE_pt}d_s^\dagger(-\vep)v_s(-\vep)\Bigr],
\label{psi}
\end{split}
\ee
with $E_p$ for the dispersion law of the dressed fermions parameterised in terms of a function of the 3-momentum $\vp(p)\equiv\vp_p$ known as the chiral angle,
\be \left\{
\begin{array}{rcl}
u(\vep)&=&\ds\frac{1}{\sqrt{2}}\left[\sqrt{1+\sin\vp_p}+
\sqrt{1-\sin\vp_p}\;(\vec{\alpha}\hat{\vep})\right]u(0)\\
v(-\vep)&=&\ds\frac{1}{\sqrt{2}}\left[\sqrt{1+\sin\vp_p}-
\sqrt{1-\sin\vp_p}\;(\vec{\alpha}\hat{\vep})\right]v(0).
\end{array}
\right.
\label{uandv}
\ee
By convention, the chiral angle varies in the range
$-\frac{\pi}{2}<\vp_p\leqslant \frac{\pi}{2}$ with the boundary
conditions $\vp(0)=\frac{\pi}{2}$, $\vp(p\to\infty)\to 0$.
The profile of the chiral angle is determined from a nonlinear integral mass-gap equation. Below we briefly outline its derivation. We start from the bare fermion propagator,
\be
S_0^{-1}(p_0,\vep)=\gamma_0 p_0-{\bm\gamma}\vep-m_q,
\label{S0T0}
\ee
and dress it by summing the Dyson series of planar rainbow diagrams as depicted in Fig.~\ref{fig:dyson},
\be
S=S_0+S_0\Sigma S_0+S_0\Sigma S_0\Sigma S_0+\ldots=S_0+S_0\Sigma S,
\label{Ds}
\ee
that gives for the dressed quark propagator
\be
S^{-1}(p_0,\vep)=S_0^{-1}(p_0,\vep)-\Sigma(\vep),
\label{Sm1}
\ee
where the self-energy operator is
\be
i\Sigma(\vep)=\int\frac{d^4k}{(2\pi)^4}V(\vep -\vek)\gamma_0 S(k_0,{\vek})\gamma_0.
\label{Sigma0}
\ee
Here $V(\vep)$ represents the confining gluon propagator discussed in detail in the next subsection --- see Eq.~\eqref{FV} below. The dressed Green's function in Eq.~\eqref{Sm1} can be conveniently decomposed into the positive- and negative-energy components,
\be
S(p_0,\vep)=\frac{\Lambda_+(\vep)\gamma_0}{p_0-E_p+i\epsilon}+
\frac{{\Lambda_-}(\vep)\gamma_0}{p_0+E_p-i\epsilon},
\label{Feynman}
\ee
where the projectors read
\be
\Lambda_\pm(\vep)=T_pP_\pm T_p^\dagger=\frac12[1\pm\gamma_0\sin\vp_p\pm({\bm\alpha}\hat{\vep })\cos\vp_p],
\label{Lpm}
\ee
with $T_p$ for the Foldy operator,
\be
T_p=\exp\left[-\frac12({\bm\gamma\hat{\vep}})\left(\vph_p-\frac{\pi}2\right)\right],
\ee
$P_\pm=\frac12(1\pm\gamma_0)$, and $\hat{\vep}$ for the unit vector in the direction of the momentum $\vep$.

Solution of the coupled equations (\ref{Sm1}) and (\ref{Sigma0}) can be found in the form
\begin{align}
&S^{-1}(p_0,\vep)=\gamma_0p_0-({{\bm\gamma}}\hat{\vep })B_p-A_p,\nonumber\\[-1mm]
\label{ST0}\\[-1mm]
&\Sigma(\vep)=[A_p-m_q]+({{\bm\gamma}\hat{\vep }})[B_p-p],
\nonumber
\end{align}
with the auxiliary functions $A_p$ and $B_p$ given by
\bea
A_p&=&\mq+\frac12\int\frac{d^3k}{(2\pi)^3}V(\vep-\vek)\sin\vp_k,\nonumber\\[-2mm]
\label{AB}\\[-2mm]
B_p&=&p+\frac12\int \frac{d^3k}{(2\pi)^3}\;
(\hat{\vep}\hat{\vek})V(\vep-\vek)\cos\vp_k\nonumber
\eea
and the chiral angle $\vp_p$ satisfying a mass-gap equation,
\be
A_p\cos\vp_p-B_p\sin\vp_p=0.
\label{mge}
\ee
Then the energy of the dressed fermion (dispersion law) introduced in Eq.~\eqref{psi} and entering Eq.~\eqref{Feynman} is calculated as
\be
E_p=A_p\sin\vp_p+B_p\cos\vp_p.
\label{Ep}
\ee

Alternatively, the mass-gap equation \eqref{mge} can be derived from the requirement that the vacuum energy is a minimum in terms of the dressed quark field (\ref{psi}) \cite{Amer:1983qa,LeYaouanc:1984ntu,Adler:1984ri} or, equivalently, that the term in the Hamiltonian (\ref{GNJL}) quadratic in the dressed quark/antiquark creation and annihilation operators does not contain anomalous (proportional to $b^\dagger d^\dagger$ and $db$) contributions \cite{Bicudo:1989sh}. The mass-gap equation \eqref{mge} with the linear confining potential \eqref{Vlin} was first solved in Ref.~\cite{Adler:1984ri}. Later, this solution was revisited in many subsequent theoretical papers. It was also generalised to other shapes of the confining potential --- see, for example, Refs.~\cite{Amer:1983qa,Bicudo:2003cy} for a detailed study of power-like confining potentials.

In the limit of a vanishing interaction, $V(\vec p)\equiv 0$, the solution of the mass-gap equation \eqref{mge} is simply
\be
\sin\vp^{(0)}_p=\frac{\mq}{\sqrt{p^2+\mq^2}},\quad
\cos\vp^{(0)}_p=\frac{p}{\sqrt{p^2+\mq^2}},
\label{vp0}
\ee
and, quite predictably, the quark dispersion law \eqref{Ep} takes its free form, $E^{(0)}_p=\sqrt{p^2+\mq^2}$. For $V(\vec p)\neq 0$, the mass-gap equation is subject to numerical investigations and, if its nontrivial solution $\vp_p$ is found, a measure of spontaneous breaking of chiral symmetry in the vacuum is provided by the chiral condensate evaluated as (for one light quark flavour)
\be
\braket{\bar{\psi}\psi}=-\frac{N_c}{\pi^2}\int^{\infty}_0 dp\;p^2\Bigl(\sin\vp_p-\sin\vp_p^{(0)}\Bigr),
\label{Sigma1}
\ee
where we subtracted the effect of the explicit symmetry breaking through the current quark mass $\mq$.

\subsection{Infrared-infinite and infrared-finite quantities}
\label{sec:ir}

The linear potential in Eq.~\eqref{Vlin} is infrared-singular and a suitable infrared regularisation is required to make the equations meaningful. Clearly, physical observables cannot depend on the adopted regularisation scheme. To proceed, here we employ the prescription in Refs.~\cite{Alkofer:2005ug,Wagenbrunn:2007ie,Bicudo:2010qp} that consists in introducing an infrared regulator to the confining ``gluon propagator'' in momentum space,
\be
V(p)=\frac{8\pi\sigma}{(p^2+\muIR^2)^2}.
\label{FV}
\ee
Then, in coordinate space, we have
\be
V(r)=\int \frac{d^3 p}{(2\pi)^3} V(p) e^{i\vec p \vec r}=\frac{\sigma}{\muIR}e^{-\muIR r}.
\label{div}
\ee

Notice that $V(r)$ in Eq.~\eqref{div} has just the opposite sign
(because of the factors $i$ that appear in two interaction vertices) to the confining potential in Eq.~\eqref{Vlin}.
Then it is easy to see that, in the limit of $\muIR\to 0$,
\be
V_{\rm conf}(r)=-V(r)\mathop{=}_{\muIR\to 0}-\frac{\sigma}{\muIR}+\sigma r+\ldots,
\label{Vreglim}
\ee
where the ellipsis stands for the terms suppressed for $\muIR\to 0$.
This way the original form of the linear confinement in Eq.~\eqref{Vlin} is readily restored, but an infrared-divergent constant appears in addition. Since the interquark potential is not observable, the presence of a singular constant in it does not cause any trouble. Moreover, all nonobservable coloured quantities acquire such singular contributions \cite{Glozman:2008fk} and are consequently removed
from the spectrum. In particular, it is easy to verify that, after regularisation with $\muIR$ \cite{Wagenbrunn:2007ie},
\begin{align}
A_p=\frac{\sigma}{2\muIR}\sin\varphi_p+A_p^{\rm fin},\nonumber\\[-3mm]
\label{ABir}\\[-3mm]
B_p=\frac{\sigma}{2\muIR}\cos\varphi_p+B_p^{\rm fin},\nonumber
\end{align}
with $A_p^{\rm fin}$ and $B_p^{\rm fin}$ for infrared-finite contributions. Then the function $E_p$ in Eq.~\eqref{Ep} is also infrared divergent,
\be
E_p=\frac{\sigma}{2\muIR}+\Bigl(A_p^{\rm fin}\sin\vp_p+B_p^{\rm fin}\cos\vp_p\Bigr).
\label{Epir}
\ee
Consequently, the single-quark Green's function (\ref{Feynman}) does not have finite physical poles in the complex energy plane, which admits a natural interpretation that quarks and antiquarks cannot be on mass shell. Meanwhile, observable quantities are free of the infrared divergences, so that the singular terms $\propto 1/\muIR$ must drops out from them. For example, it is easy to see that this term cancels in the mass-gap equation (\ref{mge}) and, therefore, the chiral angle $\varphi_p$ is also finite in the infrared limit. The same holds for the observable colour-singlet meson masses since the infrared divergence of the single quark Green's function is exactly cancelled by the infrared divergence of the kernel in the Bethe--Salpeter equation for a quark--antiquark bound state \cite{Wagenbrunn:2007ie}.

Infrared singularity of the fermion dispersion law \eqref{Epir} implies that the latter cannot be used in a suitable definition of an infrared-finite effective quark mass. Instead, it proves convenient to introduce an infrared-finite quark energy $\omega_p$ through the relation
\be
B_p\omega_p=pE_p
\label{omegapdef}
\ee
that, for a free quark, trivially yields
\be
\omega_p^{(0)}=E_p^{(0)}=\sqrt{p^2+\mq^2}.
\ee
In the meantime,
$\omega_p$ is finite in the infrared limit of $\muIR\to 0$ also for interacting quarks,
\be
\omega_p=p\lim_{\muIR\to 0}\frac{E_p}{B_p}=
\frac{p}{\cos\varphi_p}=\sqrt{p^2+(p\tan\varphi_p)^2},
\ee
so an infrared-finite dynamical quark mass can be conveniently defined as
\be
M_p=p\tan\varphi_p,
\label{Mp}
\ee
and, consequently,
\be
\sin\varphi_p=\frac{M_p}{\sqrt{p^2+M_p^2}},\quad \cos\varphi_p=\frac{p}{\sqrt{p^2+M_p^2}}.
\label{scp}
\ee
In the no-interaction limit, Eq.~\eqref{vp0} implies that $M_p^{(0)}=m_q$ while a nonvanishing interaction provides an additional contribution to $M_p$. Since the chiral angle is infrared-finite then, obviously, so is $M_p$. Indeed, a smooth $\muIR\to 0$ limit was established numerically for both the chiral angle $\vp_p$ and the effective quark mass $M_p$ in Refs.~\cite{Wagenbrunn:2007ie,Bicudo:2010qp}. More generally, from now onwards all quantities are regularised in the infrared and only those of them are studied that possess a well-defined infrared limit.

\subsection{Mass-gap equation at $T>0$}
\label{sec:mgeT}

Below we derive the mass-gap equation at finite temperatures
employing a real time formalism following Ref.~\cite{Glozman:2024xll}. The derivation proceeds along the same lines as explained in Subsec.~\ref{sec:SBCS} above, with the only exception that now, unlike at $T=0$, the thermal averages $\braket{b_s^\dagger(\vep)b_{s}(\vep)}=n_p$ and
$\braket{d_s^\dagger(\vep)d_{s}(\vep)}=\bar{n}_p$ do not vanish any more. Instead, due to the presence of quark--antiquark pairs in the thermodynamical ensemble at $T>0$, they take nonzero values,
\begin{align}
n_p=\left(1 + e^{(\sqrt{p^2 + M_p^2}-\mu)/T}\right)^{-1},\nonumber\\[-2mm]
\label{nnnew}\\[-2mm]
\bar{n}_p=\left(1 + e^{(\sqrt{p^2 + M_p^2}+\mu)/T}\right)^{-1},\nonumber
\end{align}
that can be regarded as an infrared-finite generalisation of the relations in Eq.~\eqref{FD} for the NJL model.

Then, it is easy to verify that the dressed quark propagator acquires additional contributions proportional to $n_p$ and $\bar{n}_p$,
\be
\begin{split}
S(p_0,\vep;T)=S(p_0,\vep;T=0)
&+2\pi i\Bigl[n_p\Lambda_+(\vep)\delta(p_0-E_p)\\
&-\bar{n}_p\Lambda_-(\vep)\delta(p_0+E_p)\Bigr]\gamma_0,
\label{ST}
\end{split}
\ee
with the zero-temperature dressed fermionic Green's function $S(p_0,\vep;T=0)\equiv S(p_0,\vep)$ defined in Eq.~(\ref{Feynman}) and the projectors $\Lambda_\pm(\vep)$ introduced in Eq.~(\ref{Lpm}). For the free chiral angle in Eq.~\eqref{vp0}, the Green's function \eqref{ST} readily turns to the one known in the literature \cite{Asakawa:1989bq},
\be
\begin{split}
&S_0(p_0,\vep;T)=(\slashed{p}-\mq)^{-1}\\
&+2\pi i(\slashed{p}+\mq)\delta(p^2-\mq^2)\Bigl[\theta(p_0)n_p^{(0)}+\theta(-p_0)\bar{n}^{(0)}_p\Bigr],
\label{S0T}
\end{split}
\ee
with
\begin{align}
n_p^{(0)}=\left(1 + e^{(\sqrt{p^2 + \mq^2}-\mu)/T}\right)^{-1},\nonumber\\[-2mm]
\label{nn0}\\[-2mm]
\bar{n}_p^{(0)}=\left(1 + e^{(\sqrt{p^2 + \mq^2}+\mu)/T}\right)^{-1}.\nonumber
\end{align}

The next steps are identical to those taken in Subsec.~\ref{sec:SBCS} and result in a generalisation of the auxiliary functions $A_p$ and $B_p$ (we now use tilde to distinguish them from those in Eq.~\eqref{AB} obtained at $T=0$),
\begin{align}
\tilde{A}_p&=\mq+\frac12\int\frac{d^3k}{(2\pi)^3}(1-n_k-\bar{n}_k)V(\vep-\vek)\sin\vp_k,\nonumber\\[-2mm]
\label{tildeAB}\\[-2mm]
\tilde{B}_p&=p+\frac12\int \frac{d^3k}{(2\pi)^3}(1-n_k-\bar{n}_k)
(\hat{\vep}\hat{\vek})V(\vep-\vek)\cos\vp_k,\nonumber
\end{align}
where the following easily verifiable relations,
\be
\begin{split}
n_p\Lambda_+(\vep)-\bar{n}_p\Lambda_-(\vep)&=
\frac12(n_p+\bar{n}_p)[\Lambda_+(\vep)-\Lambda_-(\vep)]\\
&+\frac12(n_p-\bar{n}_p)[\Lambda_+(\vep)+\Lambda_-(\vep)]
\label{rels}
\end{split}
\ee
and
\be
\begin{split}
&\Lambda_+(\vep)+\Lambda_-(\vep)=\gamma_0,\\
&\Lambda_+(\vep)-\Lambda_-(\vep)=\sin\vp_p+({\bm\gamma}\hat{\vep})\cos\vp_p,
\end{split}
\ee
were used. It is then easy to see that, for a nonvanishing chemical potential $\mu\neq 0$, the second term on the right-hand side in Eq.~\eqref{rels} generates an additional structure in the mass operator defined in Eq.~\eqref{ST0} that is proportional to the Dirac matrix $\gamma_0$,
\be
\Sigma(\vep ;T)=[\tilde{A}_p-m]+(\boldsymbol {\gamma}\hat{\vep })[\tilde{B}_p-p]+\gamma_0\tilde{C}_p,
\label{SigmaT}
\ee
with
\be
\tilde{C}_p=-\frac12\int\frac{d^3k}{(2\pi)^3}(n_k-\bar{n}_k)V(\vep-\vek).
\label{tildeC}
\ee
In the absence of the chemical potential ($\mu=0$), $\bar{n}_p=n_p$ and, therefore, $\tilde{C}_p=0$. Then the thermal infrared-finite mass-gap equation \eqref{mge} takes the form
\be
\tilde{A}_p\cos\vp_p-\tilde{B}_p\sin\vp_p=0.
\label{mgeT}
\ee
We notice that alternatively the expressions in Eqs.~\eqref{tildeAB} and \eqref{tildeC} can be derived employing the imaginary time formalism \cite{Kocic:1985uq}.
We also point out that using infrared-divergent dispersion law $E_p$ instead of $\sqrt{p^2 + M_p^2}$ in the distribution functions in Eq.~\eqref{nnnew} would lead to an infrared divergence in the thermal mass-gap equation \eqref{mgeT}, and the notion of the temperature would be ill-defined.

\begin{figure}[t!]
\centering
\includegraphics[width=\columnwidth]{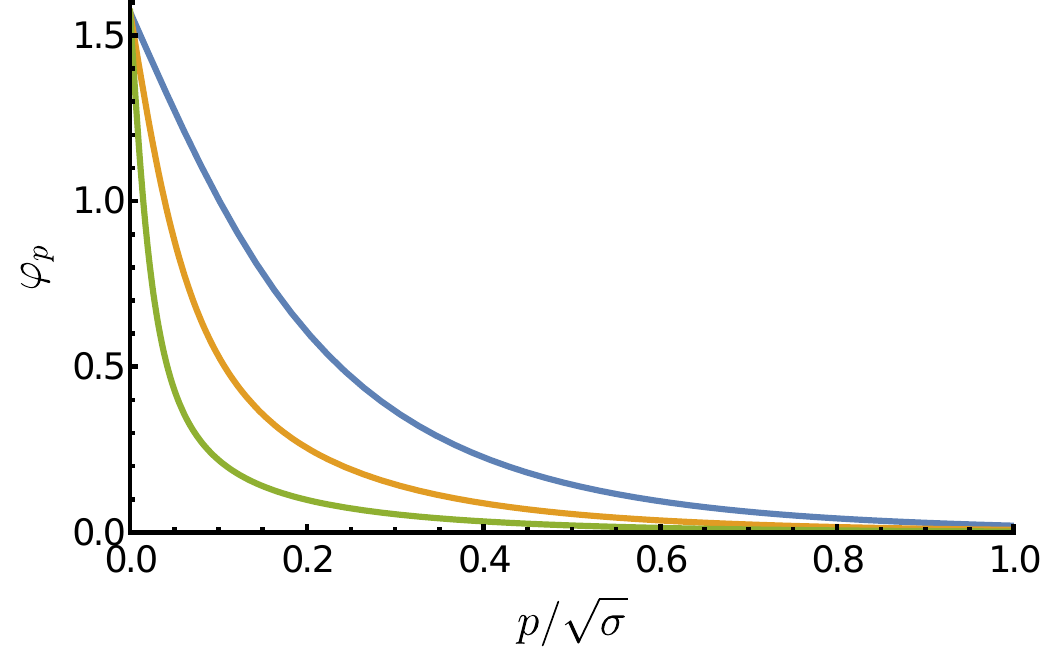}
\includegraphics[width=\columnwidth]{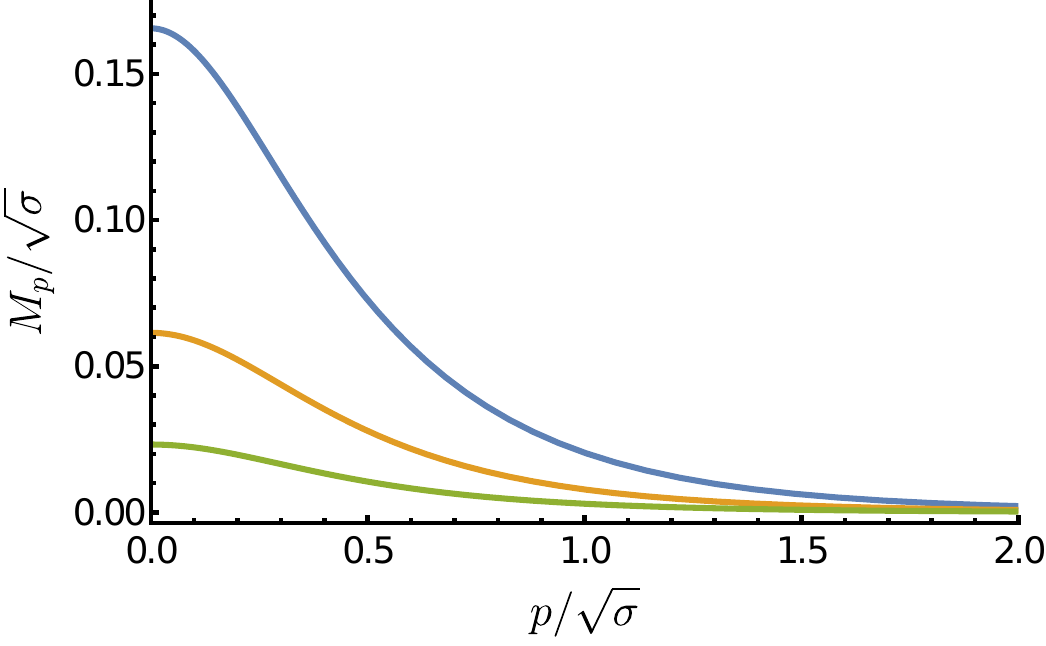}
\caption{Solution of the thermal mass-gap equation \eqref{mgeT} for the chiral angle $\vp_p$ (left plot) and the effective dressed quark mass $M_p$ in Eq.~\eqref{Mp} (right plot) for $T=0$ (upper blue curve), $T=0.95\Tch$ (middle orange curve), and $T=0.99\Tch$ (lower green curve); the value of $\Tch$ is quoted in Eq.~\eqref{Tch}.}
\label{fig:phiMp}
\end{figure}

\begin{figure}[t!]
\centering
\includegraphics[width=\columnwidth]{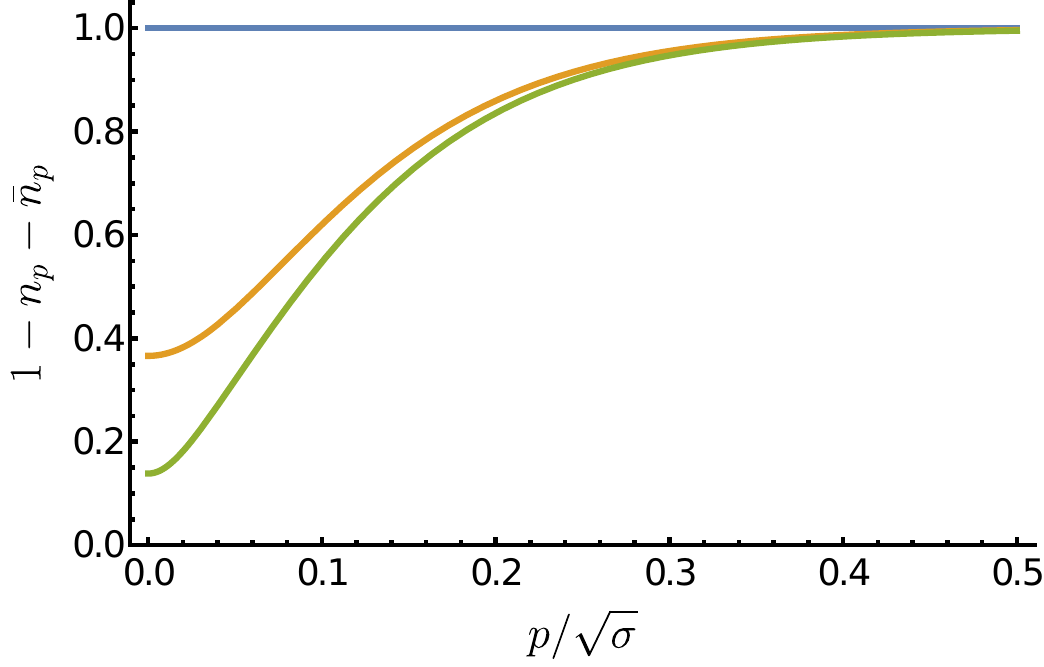}
\caption{Temperature dependence of the damping factor $1-n_p-\bar{n}_p$ in Eq.~\eqref{tildeAB} for a vanishing chemical potential ($\mu=0$) and $T=0$ (upper blue curve),
$T=0.95\Tch$ (middle yellow curve), and
$T=0.99\Tch$ (lower green curve); the value of $\Tch$ is quoted in Eq.~\eqref{Tch}.}
\label{fig:suppression}
\end{figure}

\subsection{Numerical study of the thermal mass-gap equation  and chiral restoration}
\label{sec:mgeTnum}

\begin{figure}[t!]
\centering
\includegraphics[width=\columnwidth]{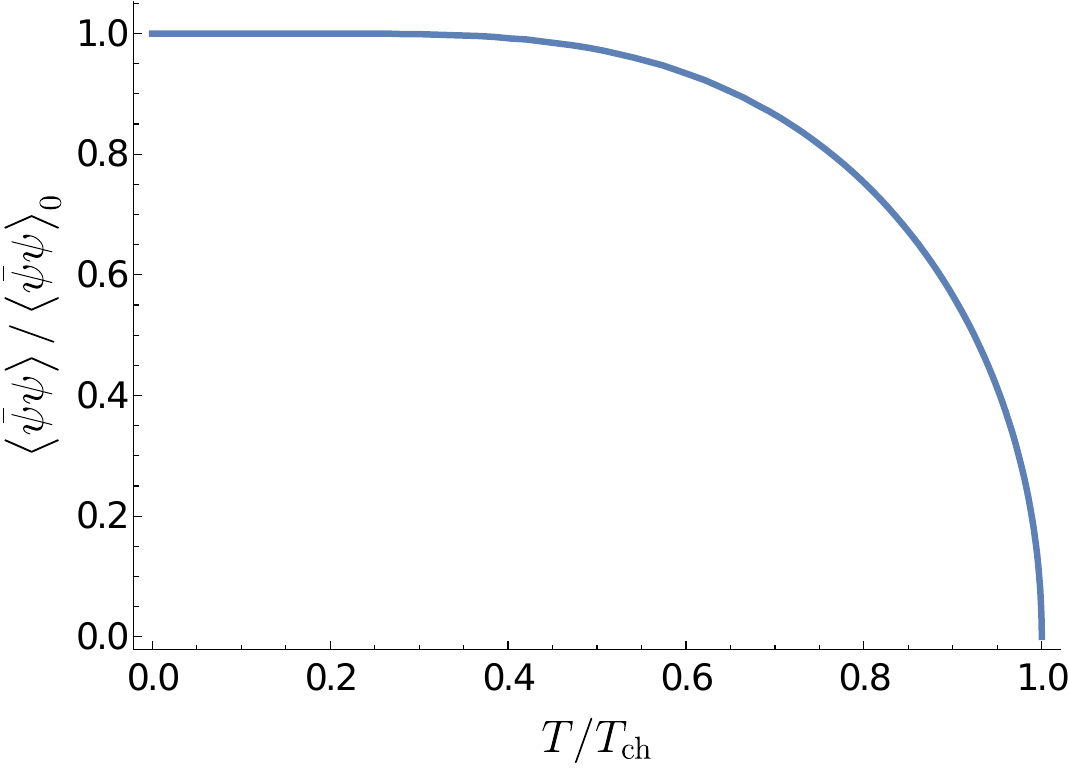}
\caption{Temperature dependence of the chiral condensate in Eq.~\eqref{Sigma1} normalised to its maximum value reached at $T=0$. Numerical values of $\braket{\bar{\psi}\psi}_0$ and $\Tch$, as they come out for the linear confining potential in Eq.~\eqref{Vlin}, are quoted in Eqs.~\eqref{cond1} and \eqref{Tch}, respectively. Adapted from Ref.~\cite{Glozman:2024xll}.}
\label{fig:cond}
\end{figure}

To proceed with numerical calculations, we stick to the linearly rising confining potential in Eq.~\eqref{Vlin}, regularised in the infrared as given in Eq.~\eqref{FV}, and work in the chiral limit of $\mq=0$. The profile of the chiral angle $\vp_p$, that comes as a solution of the thermal mass-gap equation \eqref{mgeT}, and the profile of the effective dressed quark mass $M_p$ in Eq.~\eqref{Mp} are shown in Fig.~\ref{fig:phiMp} for several values of the temperature $T$ \cite{Glozman:2024xll}. From this figure, one can conclude that, as the temperature increases, the chiral angle and the effective mass approach their free limits $\vp_p^{(0)}=0$ and $M_p^{(0)}=0$, respectively. This effect can be easily understood since the temperature-dependent factor $1-n_k-\bar{n}_k$ plotted in Fig.~\ref{fig:suppression} damps the integrals on the right-hand side of both expressions in Eq.~\eqref{tildeAB}. Importantly, the strongest suppression takes place for small momenta that are most relevant for the effect of SBCS. In the extreme case of $T\to\infty$,
the thermal Fermi-Dirac distributions approach their asymptotes, $n_p(T\to\infty)=\bar{n}_p(T\to\infty)=\frac12$, for all momenta, so the integral terms on the right-hand side in Eq.~\eqref{tildeAB} vanish and, therefore, the mass-gap equation \eqref{mgeT} possesses only a trivial solution $\vph_p\equiv 0$. From this consideration one concludes that there exists a finite critical temperature $\Tch$ such that, for $T<\Tch$, chiral symmetry is spontaneously broken in the vacuum while, for $T>\Tch$, the chiral angle identically vanishes and so does the chiral condensate in Eq.~\eqref{Sigma1}. Indeed, the stable in the limit $\muIR\to 0$ result for the temperature dependence of the chiral condensate in Eq.~\eqref{Sigma1} found in Ref.~\cite{Glozman:2024xll} is shown in Fig.~\ref{fig:cond}. In agreement with the qualitative arguments presented above, the chiral condensate $\braket{\bar{\psi}\psi}$ takes its maximal value at $T=0$,
\be
\braket{\bar{\psi}\psi}_0\equiv \braket{\bar{\psi}\psi}_{T=0}\approx-(0.23\sqrt{\sigma})^3,
\label{cond1}
\ee
and then decreases with the rise of the temperature until it finally vanishes at the critical temperature,
\be
\Tch\approx 0.084\sqrt{\sigma}.
\label{Tch}
\ee
If the string tension parameter $\sigma$ is excluded from Eqs.~\eqref{cond1} and \eqref{Tch}, the parameter-free relation predicted by the model with linear confinement reads~\cite{Glozman:2024xll}
\be
|\braket{\bar{\psi}\psi}_0|^{1/3}\approx 2.75\;\Tch.
\ee
Thus, for the phenomenological value of the chiral condensate $\braket{\bar{\psi}\psi}_0=-(250~\mbox{MeV})^3$, it predicts
\be
\Tch\approx 90~\mbox{MeV}
\ee
and this way provides a decent estimate for the lattice chiral phase transition temperature, $\Tch \simeq 130$ MeV, given a very simple form of the interquark interaction employed in the calculation.\footnote{For a more realistic approximation one needs to augment the linear potential with further terms, for example, a properly regularised colour-Coulomb interaction.}
A fit of the form,
\be
\braket{\bar{\psi}\psi}_{T\to\Tch}/\braket{\bar{\psi}\psi}_0
=c_1(1-T/\Tch)^{c_2},
\ee
performed for $\Tch-T\ll\Tch$ for the curve in Fig.~\ref{fig:cond} returns the values
\be
c_1\approx 2.39,\quad c_2\approx 0.54,
\ee
so the behaviour of the chiral condensate in approach to the critical temperature $\Tch$ from below is consistent with the dependence
\be
\braket{\bar{\psi}\psi}_{T\to\Tch}/\braket{\bar{\psi}\psi}_0\propto \sqrt{1-T/\Tch}
\ee
that is natural for the phase transition order parameter.

\subsection{Bound state equation for quark--antiquark mesons at finite $T$}
\label{sec:BS}

\begin{figure}[t!]
\centering
\includegraphics[width=0.8\columnwidth]{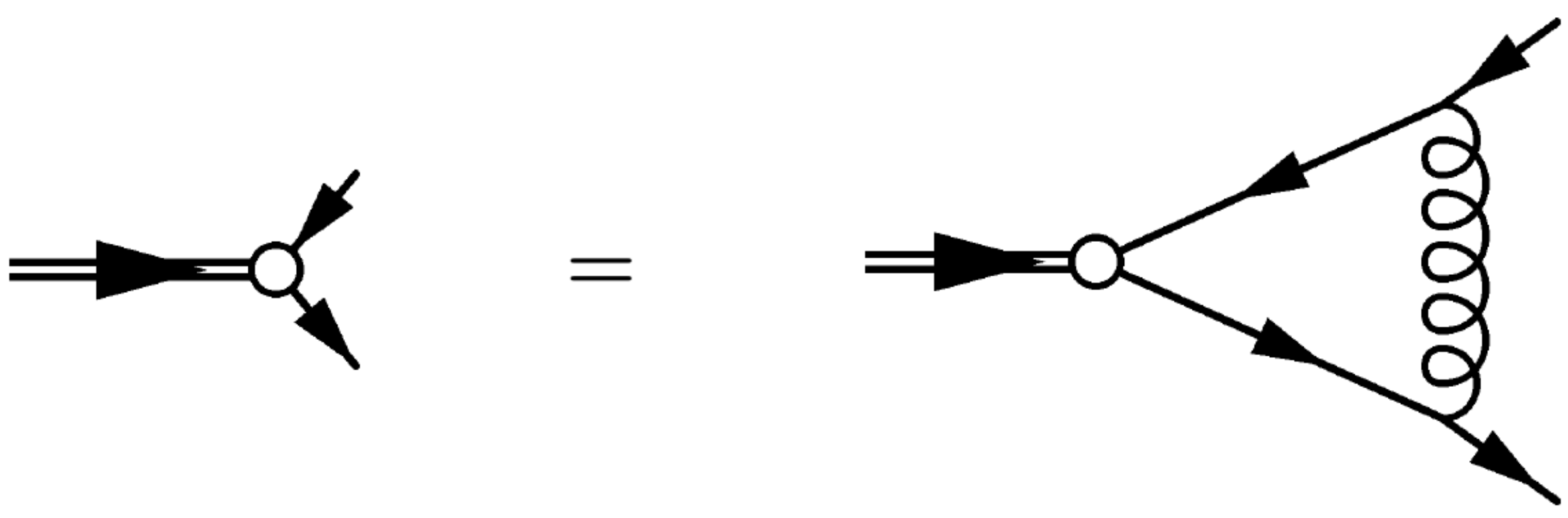}
\caption{Schematic representation of the Bethe--Salpeter equation \eqref{BSeq} written in the ladder approximation for the interquark interaction. The single, double, and curly line correspond to the quark (antiquark), meson, and the interquark interaction, respectively.}
\label{fig:bseq}
\end{figure}

The spectrum of quark--antiquark mesons can be found from the Bethe--Salpeter bound-state equation that was previously studied in many works (see, for example, Refs.~\cite{LeYaouanc:1984ntu,Bicudo:1989si,Nefediev:2004by,Wagenbrunn:2007ie}). Here we generalise it to a finite temperature $T$ first. To this end we remind the reader that a generic quark--antiquark meson in its rest
frame is described by the matrix amplitude $\chi(\vep;M)$, with $\vep$ for the momentum of the quark ($-\vep$ for the antiquark) and $M$ for the mass of the meson. The Bethe--Salpeter equation for this amplitude schematically depicted in Fig.~\ref{fig:bseq} reads
\be
\begin{split}
\chi({\vec p};M)=i\int&\frac{d^4q}{(2\pi)^4}V(\vep-\veq)\gamma_0 S(q_0+M/2,\veq;T)\\
&\times\chi(\veq;M)S(q_0-M/2,\veq;T)\gamma_0,
\label{BSeq}
\end{split}
\ee
where the propagator $S(p_0,\vep;T)$ quoted in Eq.~\eqref{ST} can be presented as a sum of the zero-temperature term and an additional temperature-dependent contribution,
\be
S(p_0,\vep;T)=S(p_0,\vep)+\Delta S(p_0,\vep;T).
\label{ST2}
\ee

Due to the instantaneous nature of the interaction described by the ``gluon propagator'' $V(\vep-\veq)$, the mesonic Bethe--Salpeter amplitude $\chi(\vep;M)$ in Eq.~\eqref{BSeq} does not depend on the temporal component of the meson momentum, so the integral in the loop energy $q_0$ can be evaluated analytically in a closed form. As a result,
the Bethe--Salpeter equation \eqref{BSeq} can be formulated as a bound-state equation for the rest-frame matrix wave function of a meson defined as
\be
\begin{split}
\matrwf(\vec{p};M)=i\int \frac{dp_0}{2\pi}
S(p_0+&M/2,\vep;T)\chi(\vep;M)\\
&\times S(p_0-M/2,\vep;T).
\label{psimatr}
\end{split}
\ee
Then, employing (i) the decomposition of the zero-temperature Green's function $S(p_0,\vep)$ into positive- and negative-energy contributions as provided in Eq.~\eqref{Feynman}, (ii) the $\delta$-functions contained in the term $\Delta S(p_0,\vep;T)$ as given in Eq.~\eqref{ST}, and (iii)
an easily verifiable relation,
\be
\begin{split}
\int\frac{dq_0}{2\pi i}\left[\frac{1}{q_0\pm
M/2-E_q+i\epsilon}\right]& \left[\frac{1}{q_0\mp
M/2+E_q-i\epsilon}\right]\\
&=-\frac{1}{2E_q\mp M},
\end{split}
\ee
we find
\be
\begin{split}
\int\frac{dq_0}{2\pi i}&S(q_0+M/2,\veq)\chi(\veq;M)S(q_0-M/2,\veq)
=\\
&-(\Lambda_+\gamma_0)\chi^{[+]}(\Lambda_-\gamma_0)
-(\Lambda_-\gamma_0)\chi^{[-]}(\Lambda_+\gamma_0),
\label{SS}
\end{split}
\ee
\begin{align}
\int&\frac{dq_0}{2\pi i}S(q_0+M/2,\veq)\chi(\veq;M)\Delta S(q_0-M/2,\veq;T)\nonumber\\
&=n_q\left[(\Lambda_-\gamma_0)\chi^{[-]}(\Lambda_+\gamma_0)
+(\Lambda_+\gamma_0)\chi^{[0]}(\Lambda_+\gamma_0)
\right]\label{SdS}\\
&+\bar{n}_q\left[(\Lambda_+\gamma_0)\chi^{[+]}(\Lambda_-\gamma_0)
-(\Lambda_-\gamma_0)\chi^{[0]}(\Lambda_-\gamma_0)\right],\nonumber
\end{align}
\begin{align}
\int&\frac{dq_0}{2\pi i}\Delta S(q_0+M/2,\veq;T)\chi(\veq;M)S(q_0-M/2,\veq)\nonumber\\
&=n_q\left[(\Lambda_+\gamma_0)\chi^{[+]}(\Lambda_-\gamma_0)
-(\Lambda_+\gamma_0)\chi^{[0]}(\Lambda_+\gamma_0)
\right]\label{dSS}\\
&+\bar{n}_q\left[(\Lambda_-\gamma_0)\chi^{[-]}(\Lambda_+\gamma_0)
+(\Lambda_-\gamma_0)\chi^{[0]}(\Lambda_-\gamma_0)\right],\nonumber
\end{align}
and
\be
\int\frac{dq_0}{2\pi i}\Delta S(q_0+M/2,\veq;T)\chi(\veq;M)\Delta S(q_0-M/2,\veq)=0,
\label{dSdS}
\ee
where
\be
\chi^{[+]}=\frac{\chi(\veq;M)}{2E_q-M},\quad
\chi^{[-]}=\frac{\chi(\veq;M)}{2E_q+M},\quad
\chi^{[0]}=\frac{\chi(\veq;M)}{M}.
\label{chipm}
\ee

Combining the results in Eqs.~\eqref{SS}-\eqref{dSdS} together, we finally arrive at the matrix wave function $\matrwf(\vec{p};M)$, defined in Eq.~\eqref{psimatr} above, in the form
\begin{multline}
\matrwf(\vec{p};M)=i(1-n_q-\bar{n}_q)\\ \times\int\frac{dq_0}{2\pi}S(q_0+M/2,\veq)\chi(\veq;M)S(q_0-M/2,\veq)\\
=(1-n_q-\bar{n}_q)\Bigl[(\Lambda_+\gamma_0)\chi^{[+]}(\Lambda_-\gamma_0)\\
+(\Lambda_-\gamma_0)\chi^{[-]}(\Lambda_+\gamma_0)\Bigr].\label{SSfin}
\end{multline}
Then the Bethe--Salpeter equation \eqref{BSeq} can be written as a system of two coupled equations,
\be
\left\{
\begin{array}{l}
[2E_p-M]\chi^{[+]}(\vep;M)=\ds\int\frac{d^3q}{(2\pi)^3}(1-n_q-\bar{n}_q)V(\vec{p}-\vec{q})\\
\times\Bigl[(\Lambda_+\gamma_0)\chi^{[+]}(\veq;M)(\Lambda_-\gamma_0)+(\Lambda_-\gamma_0)\chi^{[-]}(\veq;M)(\Lambda_+\gamma_0)\Bigr]\\[3mm]
[2E_p+M]\chi^{[-]}(\vep;M)=\ds\int\frac{d^3q}{(2\pi)^3}(1-n_q-\bar{n}_q)V(\vec{p}-\vec{q})\\
\times\Bigl[(\Lambda_+\gamma_0)\chi^{[+]}(\veq;M)(\Lambda_-\gamma_0)+(\Lambda_-\gamma_0)\chi^{[-]}(\veq;M)(\Lambda_+\gamma_0)\Bigr],
\end{array}
\label{BSeq2}
\right.
\ee
where, like in the mass-gap equation \eqref{mgeT}, the only explicit dependence on the temperature is contained in the factor $1-n_q-\bar{n}_q$. Therefore, introducing a finite temperature into the bound state equation derived at $T=0$ amounts to a replacement
$V(\vec p - \vec q) \to (1-n_q-\bar{n}_q)V(\vec p - \vec q)$, where the Fermi-Dirac distribution functions contain the dynamical quark mass $M(q) \equiv M_q$, as given in Eq.~\eqref{nnnew}, obtained from the solution of the thermal mass-gap equation at the given finite temperature $T$.
It implies that, at all temperatures, the effective interaction in the quark--antiquark meson is exactly the same as in the mass-gap equation. In particular, it guarantees that the bound state equation for the pion is equivalent to the mass-gap equation (a dualism of the pion) at all temperatures $T<\Tch$ when the pion plays a role of the Goldstone boson related to SBCS \cite{Nambu:1960xd,Goldstone:1961eq}.

To proceed, it proves convenient to define the positive- and negative-energy wave functions as
\be
\matrwf_\pm(\vep;M)=P_\pm T_p\chi^{[\pm]}(\vep;M)T_pP_\mp,
\label{psipm}
\ee
where the Foldy operator $T_p$ and projectors $P_\pm$ were introduced in Eq.~\eqref{Lpm} above.
Then the Bethe--Salpeter equation \eqref{BSeq2} can be written
in the form
\be
\left\{
\begin{array}{l}
[2E_p-M]\matrwf_+(\vep;M)=\ds\int\frac{d^3q}{(2\pi)^3}(1-n_q-\bar{n}_q)V(\vep-\veq)\\
\hspace*{0.16\columnwidth}\times\Bigl[{\cal P}_{++}\matrwf_+(\veq;M){\cal P}_{--}+{\cal P}_{+-}\matrwf_-(\veq;M){\cal P}_{+-}\Bigr]\\[3mm]
[2E_p+M]\matrwf_-(\vep;M)=\ds\int\frac{d^3q}{(2\pi)^3}(1-n_q-\bar{n}_q)V(\vep-\veq)\\
\hspace*{0.16\columnwidth}\times\Bigl[{\cal P}_{-+}\matrwf_+(\veq;M){\cal P}_{-+}+{\cal P}_{--}\matrwf_-(\veq;M){\cal P}_{++}\Bigr],
\end{array}
\label{BSeq3}
\right.
\ee
with
$$
{\cal P}_{\lambda_1\lambda_2}=P_{\lambda_1} T_pT_q^\dagger P_{\lambda_2},\quad \lambda_{1,2}=\pm.
$$
Depending on the quantum numbers of the meson, the spin and angular variables can be properly disentangled in $\matrwf_\pm(\vep;M)$ and two scalar radial wave functions,
$\psi_\pm(p)$, can be introduced to describe the forward and backward in time propagation of the quark-antiquark pair in the meson, respectively\footnote{We recall that the instantaneous interquark interaction employed in the model entails that the quark and antiquark in a meson can only move forward or backward in time in unison.} \cite{Bicudo:1989si,Nefediev:2004by,Wagenbrunn:2007ie}.
For example, for the pseudoscalar and scalar mesons, we have
\begin{align}
\Bigl[\matrwf_\pm(\vep;M)\Bigr]_{J^{PC}=0^{-+}}&=\frac{i}{\sqrt{2}}\sigma_2 Y_{00}(\hat{\vep})\psi_\pm^\pi(p),\nonumber\\[-3mm]
\label{wfs}\\[-3mm]
\Bigl[\matrwf_\pm(\vep;M)\Bigr]_{J^{PC}=0^{++}}&=\frac{i}{\sqrt{2}}({\bm\sigma}\hat{{\bm p}})\sigma_2 Y_{00}(\hat{\vep})\psi_\pm^\sigma(p),\nonumber
\end{align}
where $\sigma_2$ and ${\bm \sigma}$ are Pauli matrices and $Y_{00}$ is the lowest spherical harmonic normalised to unity, $Y_{00}=1/\sqrt{4\pi}$. We skip further details of solving the Bethe--Salpeter equation numerically and refer the reader to a pedagogical description of the corresponding procedures provided in Ref.~\cite{Wagenbrunn:2007ie}, where the zero-temperature Bethe--Salpeter equations for the mesons with all allowed quantum numbers are provided explicitly in Sec.~V. We refrain from quoting them here.

\begin{table*}[t]
\centering
\caption{A complete set of possible chiral multiplets (here $k=1,2,\ldots$) for the mesons with different values of the total spin $J$ provided in the form $I,J^{PC}$, with $I$, $P$, and $C$ for the isospin, spatial, and charge parity of the state, respectively \cite{Glozman:2007ek}. The symbol $\longleftrightarrow$ indicates that both given states belong to the same representation and as such must be degenerate. }
\label{syms}
\begin{tabular}{|c|c|c|c|c|}
\hline
$J$ & $(0,0)$ & $(1/2,1/2)_a$ & $(1/2,1/2)_b$ & $(0,1) \oplus (1,0)$\\
\hline
0 &--- & $1,0^{-+} \longleftrightarrow 0,0^{++}$ & $1,0^{++} \longleftrightarrow 0,0^{-+}$ &---\\
$2k$ & $0,J^{--}\quad 0,J^{++}$ & $1,J^{-+} \longleftrightarrow 0,J^{++}$& $1,J^{++} \longleftrightarrow 0,J^{-+}$ & $1,J^{++} \longleftrightarrow 1,J^{--}$\\
$2k-1$ & $0,J^{++}\quad 0,J^{--}$&$1,J^{+-} \longleftrightarrow 0,J^{--}$ &$1,J^{--} \longleftrightarrow 0,J^{+-}$ &$1,J^{--} \longleftrightarrow 1,J^{++}$\\
\hline
\end{tabular}
\end{table*}

\subsection{Properties of quark-antiquark mesons}

In this subsection, we provide the results of a numerical study of the Bethe--Salpeter equation \eqref{BSeq}, with the chiral angle $\vp_p$ previously found as solution to the thermal mass-gap equation \eqref{mgeT} (if not stated explicitly, the chiral limit of $\mq\to 0$ is implied). Several examples of the corresponding solutions for $M_p$ are given in Fig.~\ref{fig:phiMp}.

When chiral symmetry is restored, the dynamical quark mass vanishes, $M_p=0$, and the Bethe--Salpeter equations for the mesons belonging to the same chiral multiplet become identical \cite{Wagenbrunn:2007ie}. A complete set of all possible chiral multiplets in the notation $I,J^{PC}$, with $I$, $J$, $P$, and $C$ for the isospin, total spin, spatial, and charge parity of the state, respectively, is given in Table~\ref{syms} --- see Ref.~\cite{Glozman:2007ek} for further details. Note that the structure
of the chiral multiplets is different for mesons with different spins $J$. In particular, for some $J^{PC}$, restored chiral symmetry implies doubling
of states. The $U(1)_A$ multiplets combine the opposite spatial parity states from the distinct $(1/2,1/2)_a$ and $(1/2,1/2)_b$ multiplets of $SU(2)_L \times SU(2)_R$.
All states sharing the same spin $J>0$ fall into an
irreducible representation of $SU(4) \times SU(4)$ with dim=16 \cite{Glozman:2022zpy}.
We notice that it is sufficient to study only isovector mesons. Indeed, since the interaction in the employed model is isospin blind, then the isoscalar mesons with a given $J^{PC}$ from the $(0,0)$, $\left(\frac{1}{2},\frac{1}{2}\right)$, and $(0,1)\oplus(1,0)$ multiplets are strictly degenerate with the corresponding isovector states with the same $J^{PC}$. We also emphasise that, in the Nambu--Goldstone mode with chiral symmetry spontaneously broken, the states with the same $I,J^{PC}$ but belonging to different chiral multiplets necessarily mix with each other. In this case, the assignment to the ``multiplets'' is based on the Bethe--Salpeter amplitude dominating the wave function --- see Ref.~\cite{Wagenbrunn:2007ie} for further details.

\begin{figure}[t!]
\centering
\includegraphics[width=\columnwidth]{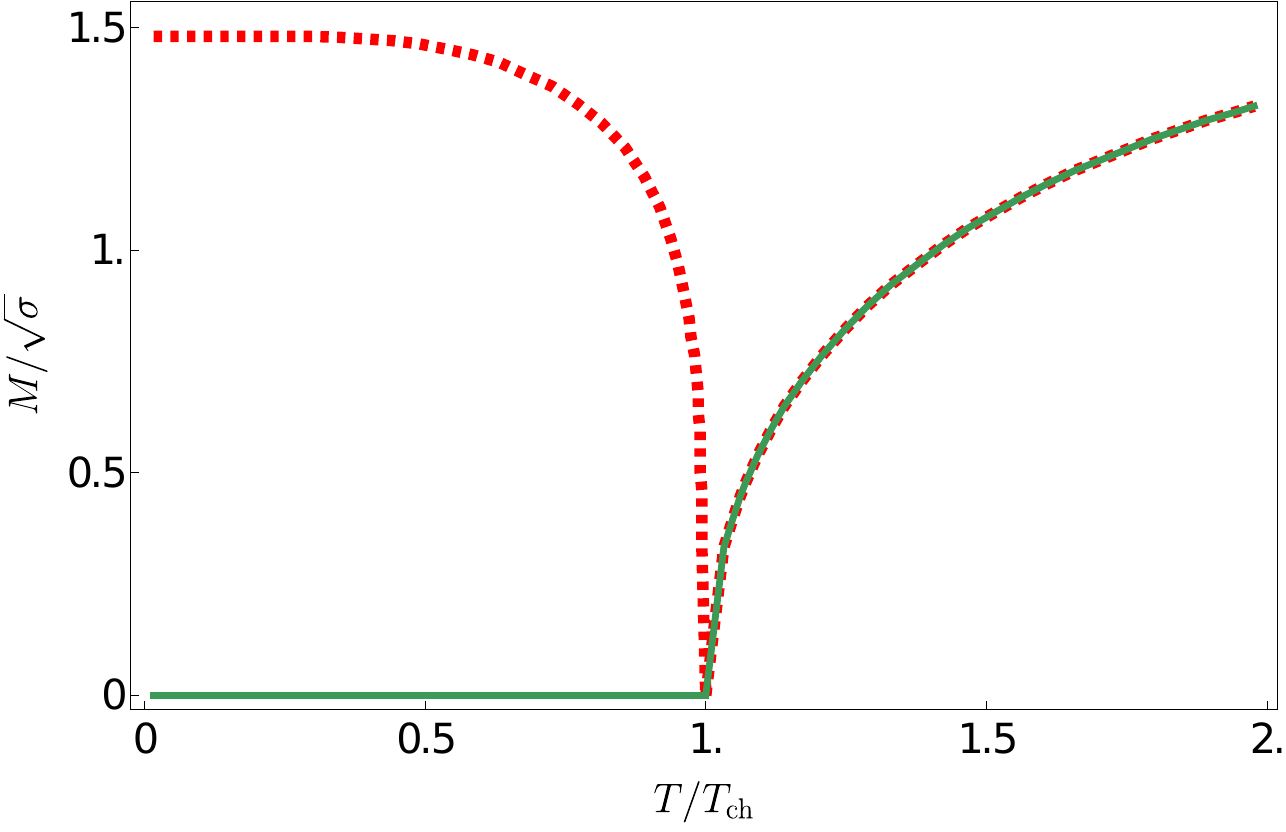}
\caption{The masses of the pseudoscalar (green solid line) and
scalar (red dotted line) mesons in the units of $\sqrt{\sigma}$ as a function of $T/\Tch$.}
\label{fig:scalarpseudoscalar}
\end{figure}

\subsubsection{The ground-state pion and scalar ``$\sigma$-meson''}

A crucial feature of the mesonic spectrum at $T<\Tch$ is a vanishing mass of the $1,0^{-+}$ ground state that is the Nambu--Goldstone boson for spontaneously broken chiral symmetry. Meanwhile, at $T>\Tch$, when chiral symmetry is restored, the dynamical quark mass vanishes, $M_p=0$, and the Bethe--Salpeter equations for the chiral pion and scalar ``$\sigma$-meson'' become identical.\footnote{Hereinafter, we use quotation marks in the name of the lowest scalar meson, solution to the Bethe--Salpater equation \eqref{BSeq3}, to distinguish it from the physical $\sigma$-meson ($f_0(500)$ \cite{ParticleDataGroup:2024cfk}) that has a more complicated structure than a plain quark--antiquark state. Indeed, while the energy-spin formalism
inherent to the chiral quark model \eqref{GNJL} naturally accounts for multiquark components of the mesons' wave functions that stem from the backward in time motion of the quarks and antiquarks (the so-called $Z$-graphs), the meson--meson interactions are automatically suppressed in the large-$N_c$ limit --- the corresponding terms in the Hamiltonian of the model scale as ${\cal O}(1/N_c)$. Then a strong $\pi$-$\pi$ interaction at low energies in the $S$ wave cannot be naturally incorporated into the given formalism and, as a result, the properties of the physical $f_0(500)$ cannot be captured properly.
This difference between the generic quark-antiquark mesons and dynamically generated objects can be well demonstrated employing the large-$N_c$ arguments. Indeed, as the number of colours grows, the mass of the generic scalar $\bar{q}q$ state remains nearly constant while its width tends to zero
since the effects of the light-quark pair creation from the vacuum scale as ${\cal O}(1/\sqrt{N_c})$ in this limit. In the meantime, the physical $f_0(500)$ pole demonstrates a severely different pattern: its real part (the mass) grows with $N_c$ while the $N_c$-dependence on its imaginary part (the width) is nontrivial, at odds with the aforementioned $1/N_C$ law specific for a generic $\bar{q}q$ scalar meson \cite{Pelaez:2006nj,Pelaez:2015qba}. It is the latter state that we denote as ``$\sigma$-meson'' and study in this work. As follows from the discussion above, at $T>\Tch$,
this scalar state must be strictly degenerate with the pseudoscalar pion.} We illustrate the pattern in Fig.~\ref{fig:scalarpseudoscalar} where we plot
the dependence of the lowest scalar and pseudoscalar meson masses on the temperature to observe that the splitting between them (i) is largest at $T=0$, (ii) remains nearly constant up to $T\approx 0.5\Tch$, and (iii) then drops fast to disappear at $T=\Tch$.

\begin{figure*}[t!]
\centering
\begin{tabular}{cc}
\includegraphics[width=0.52\textwidth]{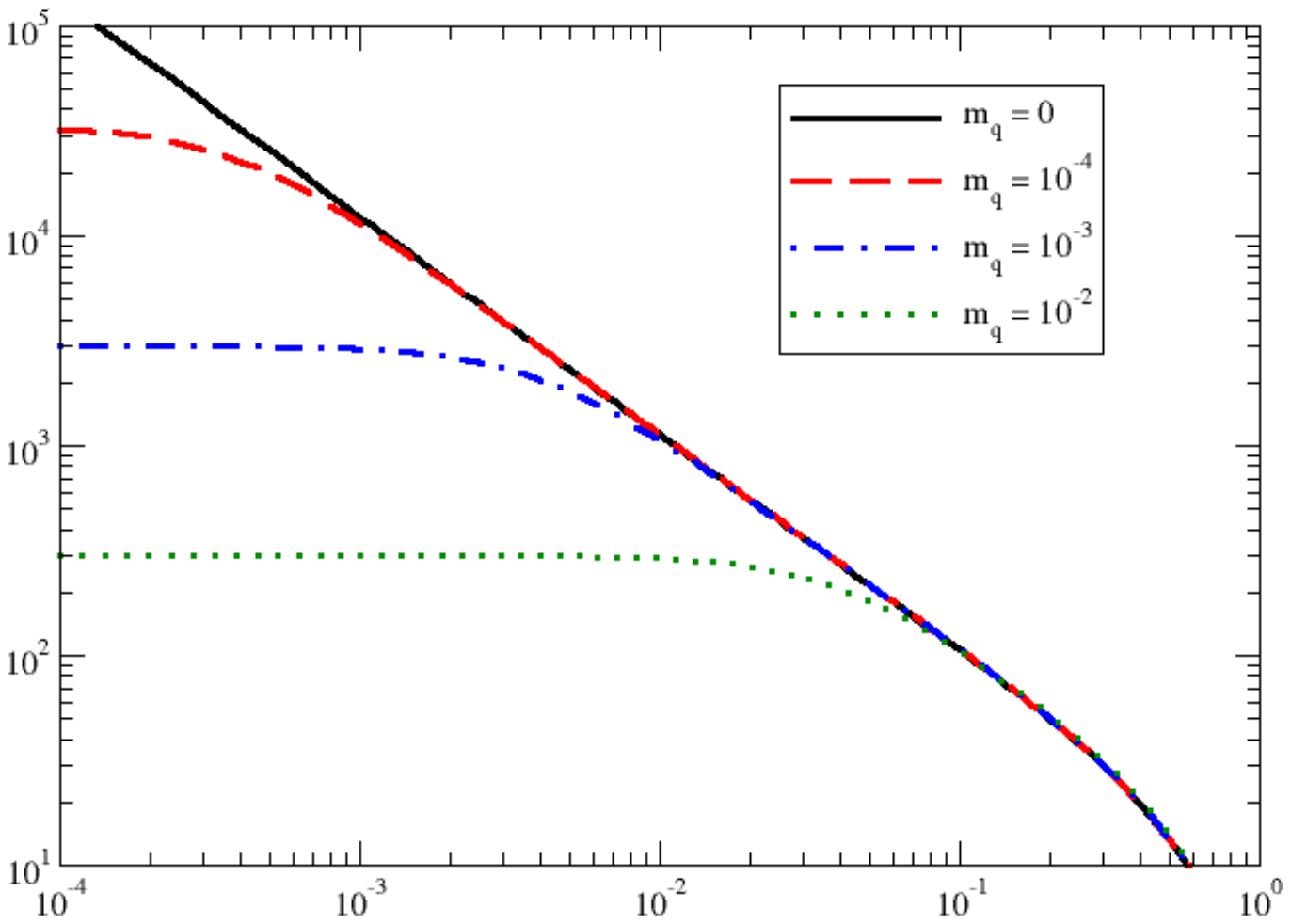}&
\hspace*{-0.06\textwidth}\includegraphics[width=0.52\textwidth]{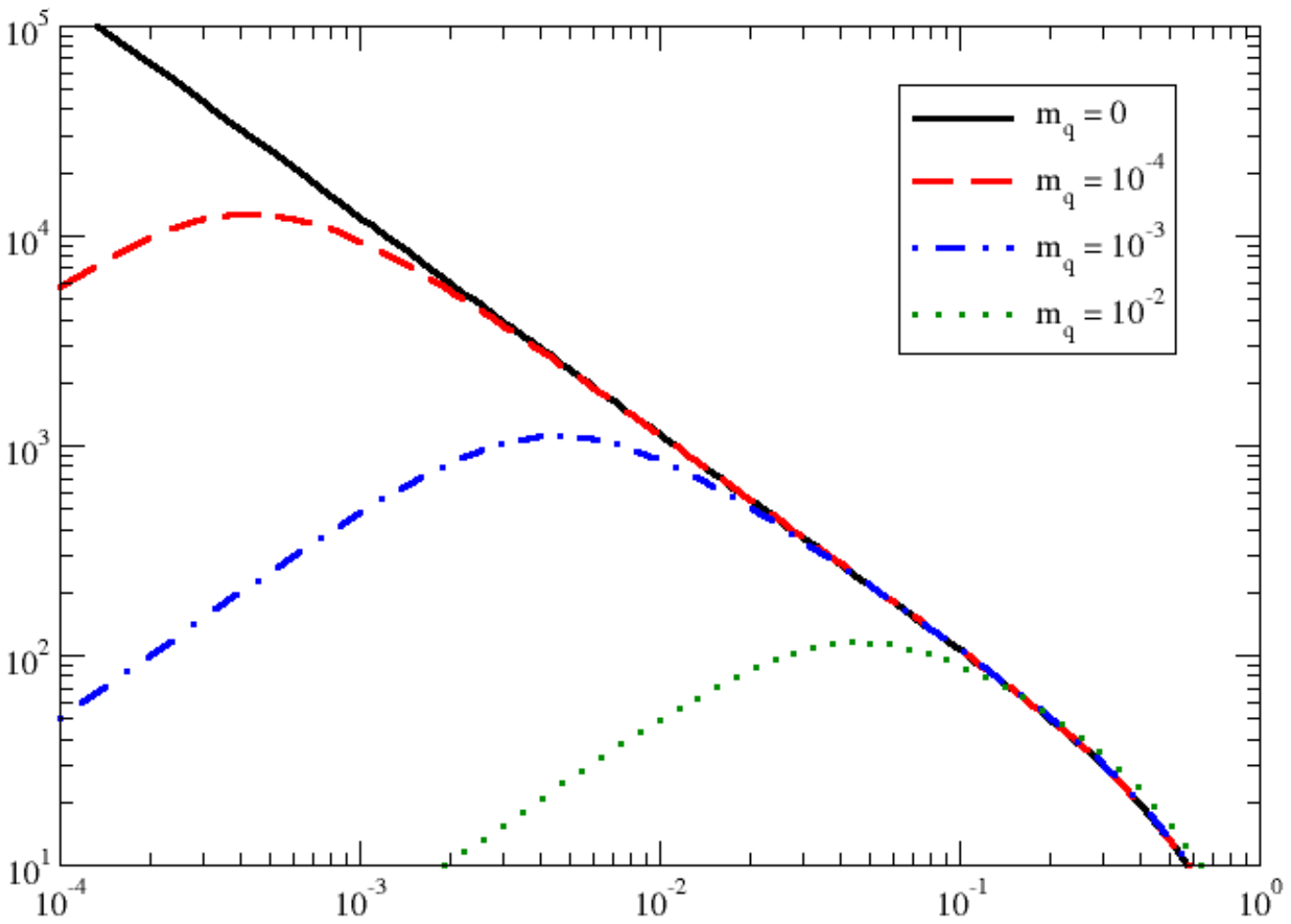}
\end{tabular}
\caption{The behaviour (in log-log scale) of the ground-state ($n=0$) wave function $\psi_+(p)$ for $J^{PC}=0^{-+}$ (left plot) and $J^{PC}=0^{++}$ (right plot) at $T=1.5\Tch$ for different small quark masses approaching the limit $m_q \to 0$. All dimensional quantities are given in the appropriate units of $\sqrt{\sigma}$.}
\label{fig:wfsdiv}
\end{figure*}

We now proceed to the wave functions $\psi_\pm(p)$ introduced in Eq.~\eqref{wfs} and to this end extend the analysis in Ref.~\cite{Wagenbrunn:2007ie} to non-zero temperatures.
When chiral symmetry is spontaneously broken at $T<\Tch$, the dynamical quark mass $M_p$ is large and the small current quark mass $\mq$ can be safely neglected from the beginning with no effect on the results. However, above $\Tch$, the chiral limit implies approaching it from the chirally broken phase with $\mq\neq 0$ and taking $\mq \to 0$ gradually.
Indeed, in QCD  the chiral limit of $\mq=0$ and the large-$N_c$ limit  do not commute.
We also remind the reader that the positive- and negative-energy wave functions for a given meson, $\psi_\pm(p)$, extracted from the Bethe--Salpeter amplitudes $\matrwf_\pm(\vep;M)$ in Eq.~\eqref{psipm} do not support a straightforward probabilistic interpretation because of their ``nonstandard'' normalisation condition,
\be
\begin{split}
\int\frac{d^3p}{(2\pi)^3}&\mbox{Tr}\Bigl[\matrwf_+^2(\vep;M)-\matrwf_-^2(\vep;M)\Bigr]\\
=&\int\frac{p^2dp}{2\pi^2}\Bigl[\psi_+^2(p)-\psi_-^2(p)\Bigr]=2M,
\label{norm2M}
\end{split}
\ee
where the trace is taken over the spin indices. The relativistic normalisation in Eq.~\eqref{norm2M} is consistent with the condition that the charge of the state with $I=I_3=1$ equals to unity \cite{Wagenbrunn:2007ie}.
In Fig.~\ref{fig:wfsdiv}, we show a log-log plot for the component $\psi_+(p)$ of the ground-state ($n=0$) pion and ``$\sigma$-meson'' wave function at $T = 1.5\Tch$ for a gradually decreasing quark mass $\mq$. We observe that the depicted radial wave functions become singular at $p=0$ when we approach the
chiral limit of $\mq \to 0$. This feature is closely related to the small-momentum behaviour of the Fermi-Dirac distributions in Eq.~\eqref{nnnew},
\be
(1-n_p-\bar{n}_p)_{|\mu=0,M_p=0,T>0}\mathop{\propto}_{p\to 0}p,
\label{PB}
\ee
that is, to the Pauli blocking effect.

\begin{figure*}[t!]
\centering
\includegraphics[width=0.9\textwidth]{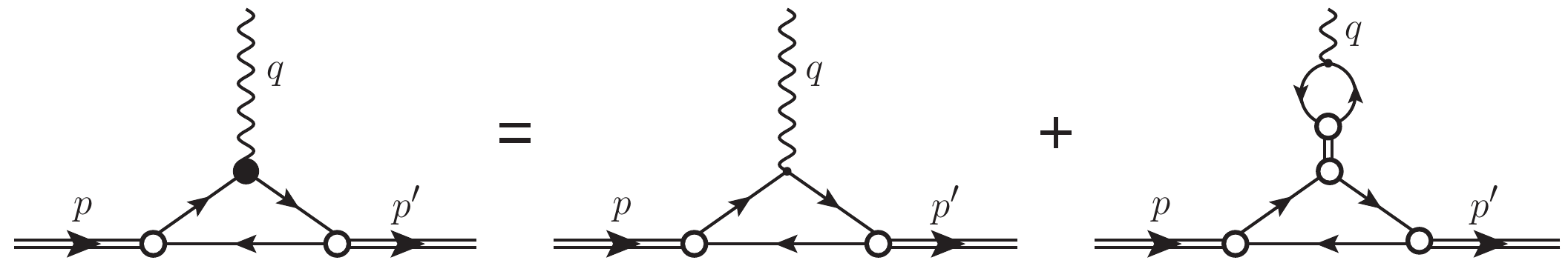}
\caption{Meson-photon interaction. The single, double, and wavy line correspond to the quark (antiquark), meson, and photon, respectively. The last diagram implies a sum over all intermediate mesons with allowed quantum numbers ($1^{--}$ vectors) coupled to the photon.}
\label{fig:mesonphoton}
\end{figure*}

\begin{figure*}[t!]
\centering
\includegraphics[width=0.49\textwidth]{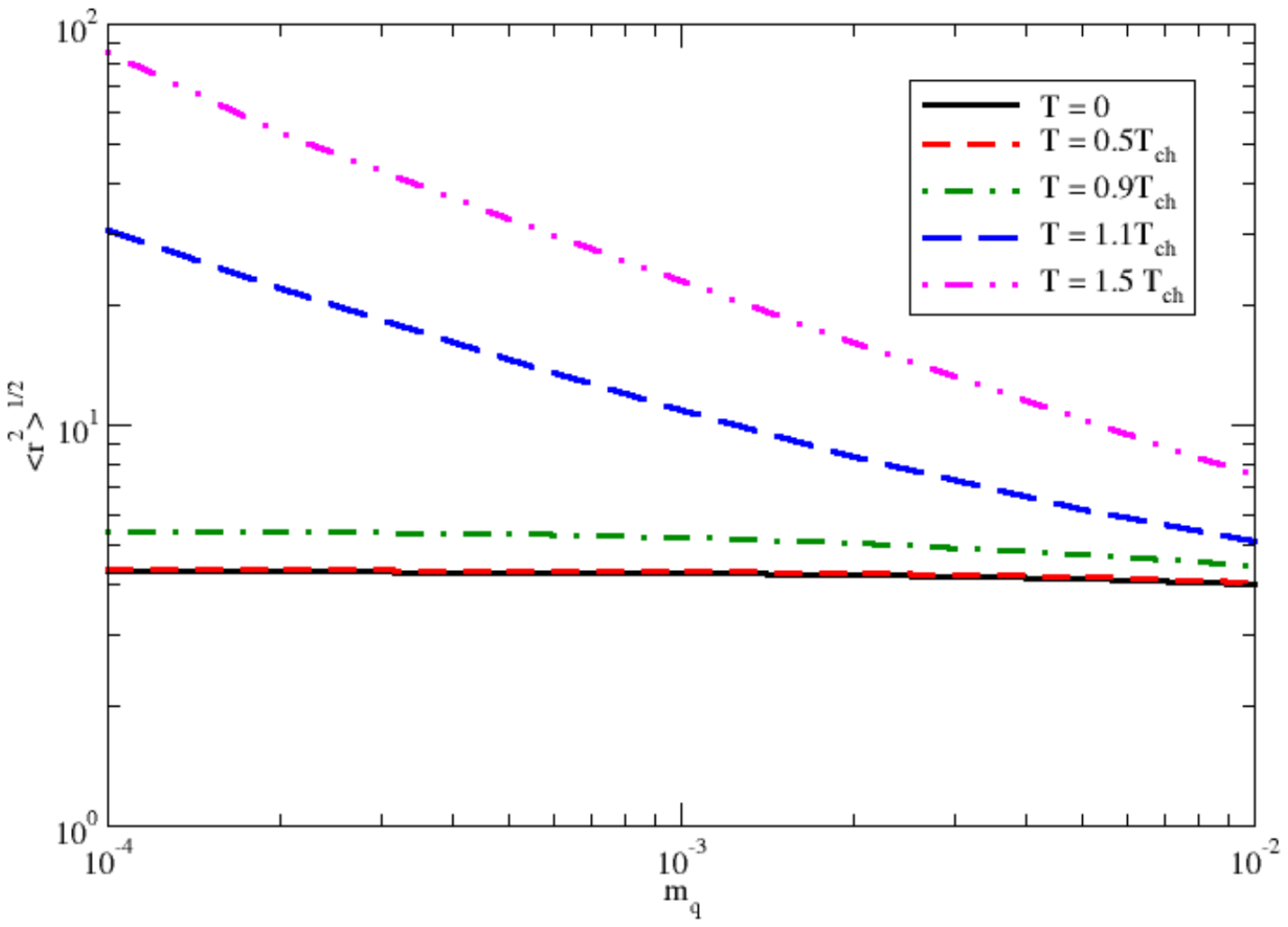}
\includegraphics[width=0.49\textwidth]{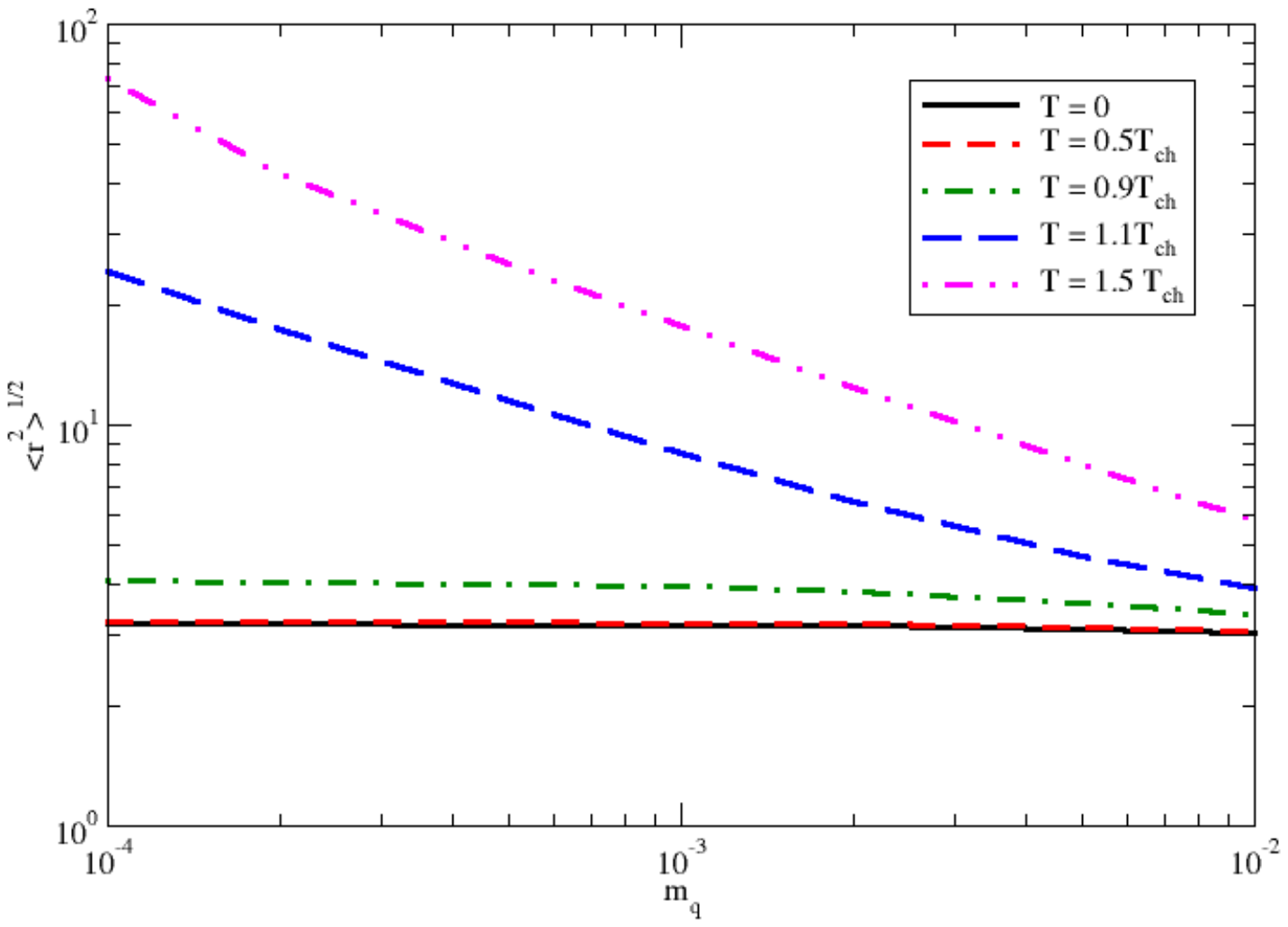}
\caption{The r.m.s. radius of the pion (first plot) and ``$\sigma$-meson'' (second plot) as function of the quark mass at different temperatures.}
\label{fig:radius}
\end{figure*}

\begin{figure}[t!]
\centering
\includegraphics[width=0.49\textwidth]{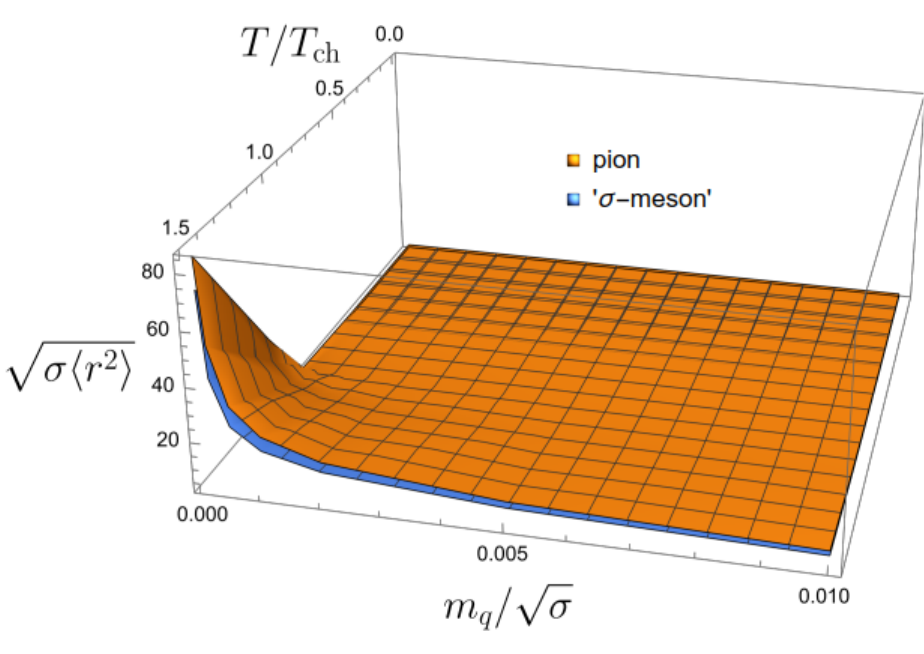}
\caption{3D plot for the r.m.s. radius of the pion (yellow) and ``$\sigma$-meson'' (blue) as function of the quark mass and temperature.}
\label{fig:radius3D}
\end{figure}

It is also instructive to check how the size of these mesons changes with the temperature. To this end, we investigate the dependence of the root-mean-square (r.m.s.) radius $\braket{r^2}^{1/2}$ \cite{Alkofer:2005ug} on $T$ and the quark mass $\mq$.
In particular, for the $\pi^+$-meson, we define
\be
\braket{r^2_\pi}=\left.6\frac{\partial F_\pi(q^2)}{\partial q^2}\right|_{q^2=0},
\label{eq:rpi}
\ee
with the charge form factor $F_\pi(q^2)$ introduced as
\be
\braket{\pi(p')|J_\mu(0)|\pi(p)}=i(p+p')_\mu F_\pi(q^2),
\label{eq:Fpi}
\ee
where $p$ and $p'$ are the pion momenta and $q=p'-p$ is the photon momentum. Conservation of the vector current in Eq.~\eqref{eq:Fpi} implies a proper dressing of the photon-quark-antiquark vertex with the interaction. In the employed quark model, this dressing amounts to the photon conversion to vector mesons via a quark--antiquark loop as depicted in Fig.~\ref{fig:mesonphoton}. Meanwhile, it follows from the definition \eqref{eq:rpi} that the main contribution to the radius comes from small momentum transfers, $q^2\to 0$, while the last diagram in Fig.~\ref{fig:mesonphoton} is operative at $q^2\gtrsim M_\rho^2$, with $M_\rho\simeq 1$~GeV for the mass of the lightest vector meson in the spectrum. Therefore, for the calculation of the r.m.s. radius we approximate the electromagnetic current by a sum of two single-quark currents (the first term on the right-hand side in Fig.~\ref{fig:mesonphoton} plus a similar term for the photon coupled to the other quark line in the loop). Since the employed formalism is not Lorentz-covariant, we use Galilean boosts for moving mesons and perform the calculations in the Breit frame. Then we proceed along the same lines for the ground-state scalar meson (we refer to it as to ``$\sigma$-meson'') as well.
We list the results obtained for the pion and ``$\sigma$-meson'' in Tables~\ref{tab:rpi} and \ref{tab:rscalar}, respectively, and visualise them in Figs.~\ref{fig:radius} and \ref{fig:radius3D}.
The value of the quark mass that would correspond to the real world is of the order $10^{-3}$ in the units of $\sqrt{\sigma}$. For this quark mass and above the chiral restoration temperature $\Tch$, the pion and ``$\sigma$-meson'' r.m.s. radii increase approximately $3\div 5$ times as compared to their radii at $T=0$. A remarkable property
of ``swelling'' of these meson-like states at finite $T$'s above $\Tch$ should also be attributed to the Pauli blocking effect. Indeed, we emphasise that, in the employed model, the confining potential is operative at all temperatures. Then, at zero and low temperatures, it guarantees that the mesons' wave functions are localised in a small space volume. However, as the temperature increases, thermal excitations of the quarks and antiquarks lead to Pauli blocking of the fermion levels with small momenta $p$ that overcomes the effect from confinement. As a result, the size of the low-lying mesons experiences an enormous (unlimited in the approach to the chiral limit\footnote{Indeed, the log-log plot in Fig.~\ref{fig:radius} suggests that, at $T>\Tch$ and for small $\mq$'s, the r.m.s radii of the pion and ``$\sigma$-meson'' scale as $\mq^{-1/2}$, as follows from the slope of the curves close to $-1/2$.}) rise as the temperature increases above $\Tch$. Nevertheless, we still deal with meson-like quark-antiquark bound states with finite well-defined masses and normalisable wave functions, as discussed above. Treating this delocalisation effect as deconfinement would be misleading given that free quarks and antiquarks do not exist, as explained in Subsec.~\ref{sec:ir} above.\footnote{Note that the discussed delocalisation takes place only for light-light and heavy-light quark-antiquark mesons, that is, for the systems involving light quarks. In heavy-heavy mesons, the temperature does not play any essential role and the Pauli blocking effect is absent. Consequently, the size of heavy-heavy mesons above $\Tch$ remains almost the same as at $T=0$.}

\begin{table*}[t!]
\centering
\caption{The r.m.s. radius $\left<r^2\right>^{1/2}$ of the pion in the units of $\sigma^{-1/2}$.}
\label{tab:rpi}
\begin{tabular}{|c|cccccccc|}\hline\hline
$T\backslash \mq$&0&$10^{-4}$&$2\cdot10^{-4}$&$5\cdot10^{-4}$&$10^{-3}$&$2\cdot10^{-3}$&$5\cdot10^{-3}$&$10^{-2}$\\\hline
0&4.31&4.31&4.31&4.30&4.28&4.25&4.16&4.03\\
$0.5\Tch$&4.36&4.35&4.35&4.34&4.32&4.29&4.19&4.05\\
$0.9\Tch$&5.46&5.44&5.42&5.35&5.26&5.11&4.78&4.45\\
$1.1\Tch$&$\infty$&30.4&21.9&14.5&10.9&8.30&6.17&5.14\\
$1.5\Tch$&$\infty$&85.0&53.7&32.6&22.8&16.1&10.3&7.49\\
\hline\hline
\end{tabular}
\caption{The r.m.s. radius $\left<r^2\right>^{1/2}$ of the ``$\sigma$-meson'' in the units of $\sigma^{-1/2}$.}
\label{tab:rscalar}
\begin{tabular}{|c|cccccccc|}\hline\hline
$T\backslash \mq$&0&$10^{-4}$&$2\cdot10^{-4}$&$5\cdot10^{-4}$&$10^{-3}$&$2\cdot10^{-3}$&$5\cdot10^{-3}$&$10^{-2}$\\\hline
0&3.19&3.19&3.18&3.18&3.17&3.15&3.10&3.03\\
$0.5\Tch$&3.23&3.23&3.22&3.22&3.21&3.19&3.13&3.05\\
$0.9\Tch$&4.10&4.08&4.07&4.02&3.95&3.84&3.61&3.38\\
$1.1\Tch$&$\infty$&24.3&17.3&11.5&8.52&6.45&4.73&3.94\\
$1.5\Tch$&$\infty$&73.2&42.0&25.2&17.6&12.5&8.01&5.80\\
\hline\hline
\end{tabular}
\end{table*}

\subsubsection{Symmetries of the spectrum of quark-antiquark mesons}

In this subsection, we discuss the symmetries of the spectrum of quark--antiquark mesons below and above the chiral restoration temperature $\Tch$. The masses of the $\bar{q}q$ bound states with different quantum numbers $nJ^{PC}$ calculated at the temperatures $T=0$, $T=0.5\Tch$, $T=0.9\Tch$, $T=1.1\Tch$, and $T=1.5\Tch$ are collected in Tables~\ref{tab:spect0}-\ref{tab:spect15tch} below. The masses quoted in Table~\ref{tab:spect0} for $T=0$ reproduce those in Ref.~\cite{Wagenbrunn:2007ie}.

\begin{figure*}[t!]
\centering
\scalebox{0.75}{\includegraphics[width=0.24\textwidth,trim={2.5cm 13.43cm 12.5cm 4.97cm}]{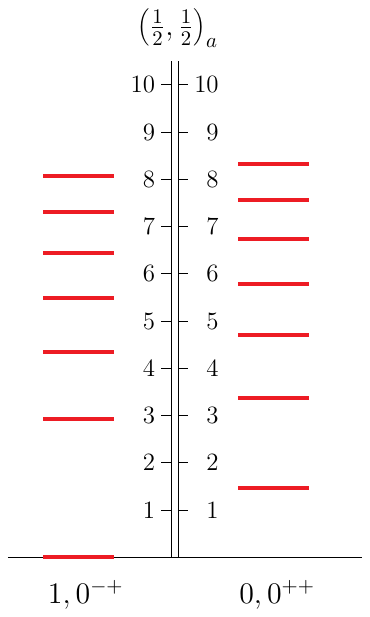}
\includegraphics[width=0.24\textwidth,trim={2.5cm 13.43cm 12.5cm 4.97cm}]{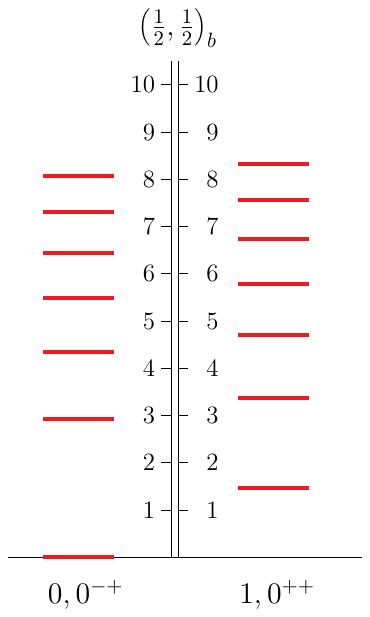}}
\caption{The spectrum (in the unites of $\sqrt{\sigma}$) of the mesons with $J=0$ at $T=0.5\Tch$.}
\label{fig:spect_05tch_j0}
\bigskip
\scalebox{0.9}{\includegraphics[width=0.24\textwidth,trim={2.5cm 13.5cm 12.5cm 5cm}]{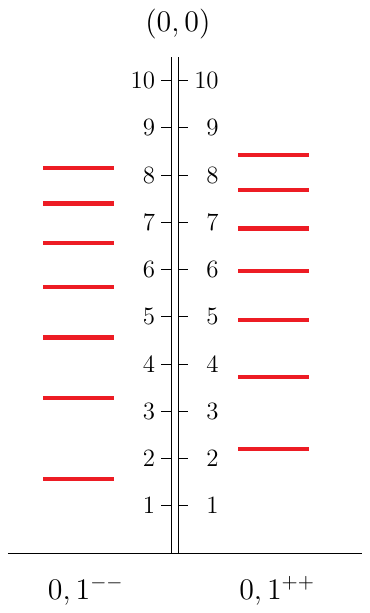}
\includegraphics[width=0.24\textwidth,trim={2.5cm 13.43cm 12.5cm 4.97cm}]{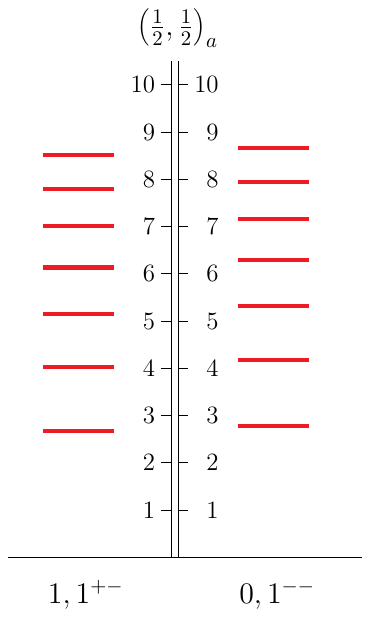}
\includegraphics[width=0.24\textwidth,trim={2.5cm 13.43cm 12.5cm 4.97cm}]{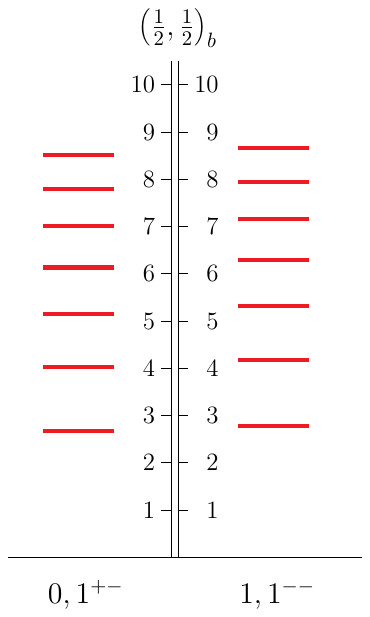}
\includegraphics[width=0.24\textwidth,trim={2.5cm 13.5cm 12.5cm 5cm}]{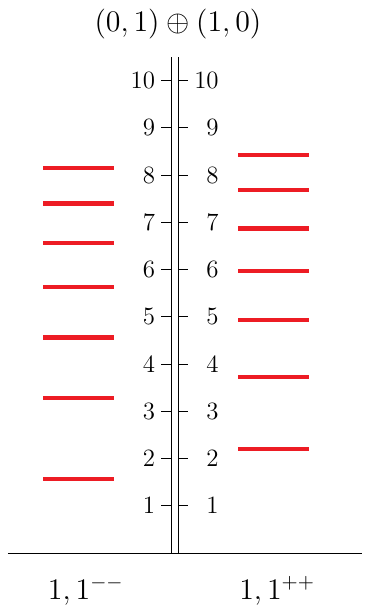}}
\caption{The spectrum (in the unites of $\sqrt{\sigma}$) of the mesons with $J=1$ at $T=0.5\Tch$.}
\label{fig:spect_05tch_j1}
\bigskip
\scalebox{0.9}{\includegraphics[width=0.24\textwidth,trim={2.5cm 13.43cm 12.5cm 4.97cm}]{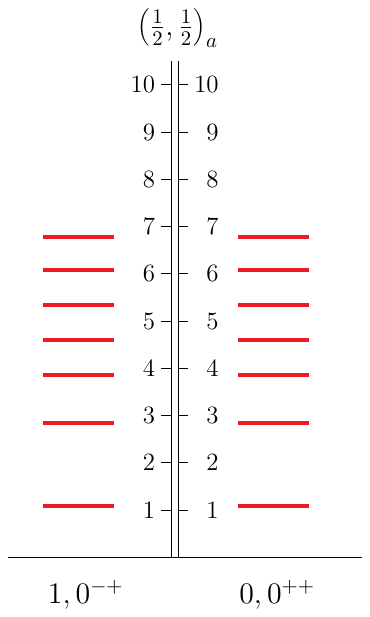}
\includegraphics[width=0.24\textwidth,trim={2.5cm 13.43cm 12.5cm 4.97cm}]{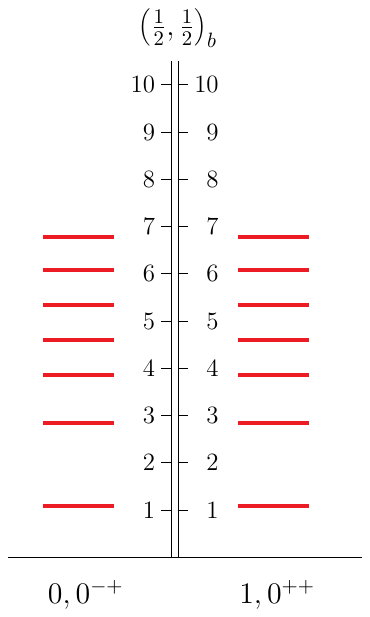}}
\caption{The spectrum (in the unites of $\sqrt{\sigma}$) of the mesons with $J=0$ at $T=1.5\Tch$.}
\label{fig:spect_15tch_j0}
\bigskip
\scalebox{0.9}{\includegraphics[width=0.24\textwidth,trim={2.5cm 13.5cm 12.5cm 5cm}]{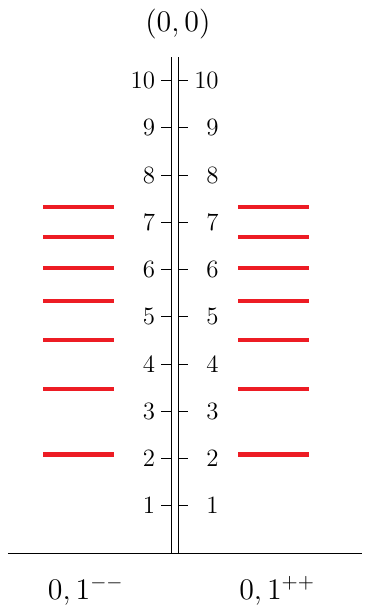}
\includegraphics[width=0.24\textwidth,trim={2.5cm 13.43cm 12.5cm 4.97cm}]{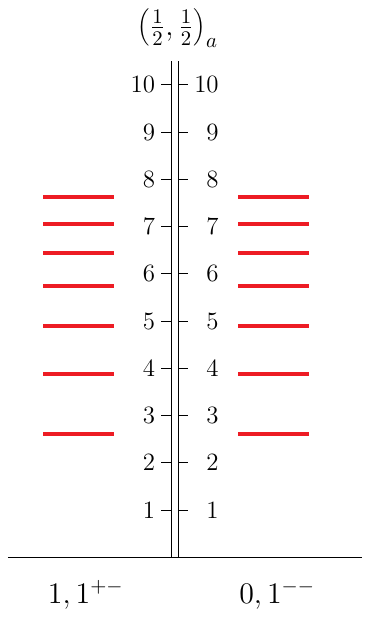}
\includegraphics[width=0.24\textwidth,trim={2.5cm 13.43cm 12.5cm 4.97cm}]{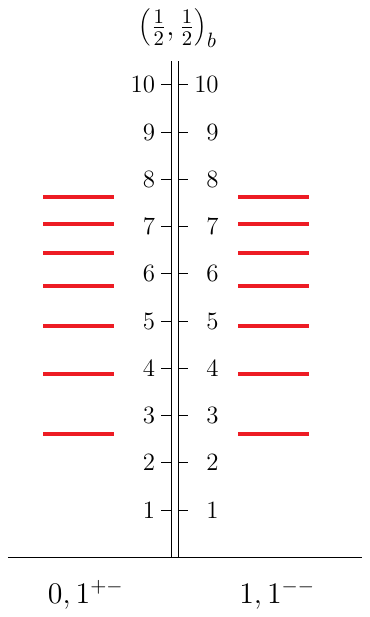}
\includegraphics[width=0.24\textwidth,trim={2.5cm 13.5cm 12.5cm 5cm}]{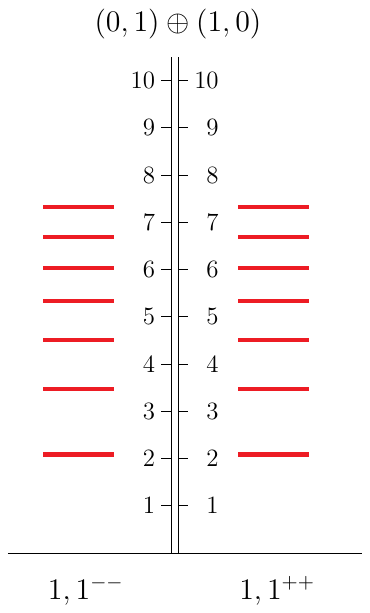}}
\caption{The spectrum (in the unites of $\sqrt{\sigma}$) of the mesons with $J=1$ at $T=1.5\Tch$.}
\label{fig:spect_15tch_j1}
\end{figure*}

At $T<\Tch$, a vanishing mass of the ground-state pion and, as a result, its large splitting with the ground-state scalar ``$\sigma$-meson'' provide a clear signal of
dynamically broken chiral symmetry as was discussed in detail in the previous subsection. Another signal of the same phenomenon is nonvanishing splittings within all chiral multiplets. As an instructive example, in Figs.~\ref{fig:spect_05tch_j0} and \ref{fig:spect_05tch_j1}, we visualise the calculated spectra of the mesonic states with $J=0$ and $J=1$ at $T=0.5\Tch$, taken from Table~\ref{tab:spect05tch}, that clearly
demonstrates the above features and this way illustrates the general pattern. We observe that the spectrum at $T=0.5\Tch$ only marginally differs from that at $T=0$, and even at $T=0.9\Tch$, noticeable changes can only be seen for the states with $J=0$ and $J=1$: the splitting within the $\left(\frac{1}{2},\frac{1}{2}\right)$ multiplets (that is, between the states with $0^{-+}$ and $0^{++}$ and between the states with $1^{+-}$ and $1^{--}$) and between the states with $1^{--}$ and $1^{++}$ belonging to the representation $(0,1)\oplus(1,0)$ diminish.

At $T>\Tch$, when chiral symmetry is restored, the dynamical quark mass vanishes, $M_p=0$, and the Bethe--Salpeter equations for the mesons belonging to the same chiral multiplet become identical \cite{Wagenbrunn:2007ie} thus resulting in the corresponding degeneracy of the spectrum. We illustrate the pattern by Figs.~\ref{fig:spect_15tch_j0} and \ref{fig:spect_15tch_j1} where, as before, we plot the spectra for the mesons with $J=0$ and $J=1$ but now at $T=1.5\Tch$ --- see Table~\ref{tab:spect15tch}.
As one can see, the splittings within the $\left(\frac{1}{2},\frac{1}{2}\right)$ and $(0,1)\oplus(1,0)$ multiplets are now completely gone, which complies well with the natural expectations given that chiral symmetry is not spontaneously broken above $\Tch$, so
the ground-state pseudoscalar meson is not the Goldstone boson any more and it becomes massive.

Furthermore, the results just obtained allow us to discuss the emergence of approximate chiral spin symmetry and its flavour extension $SU(4) \times SU(4)$ that are the symmetries of the confining part of the Hamiltonian \cite{Glozman:2022zpy}. These symmetries
are broken by the quark condensate and quark kinetic term. Then, as chiral symmetry is restored and the quark condensate vanishes, the only remaining source of the symmetry breaking is the quark kinetic term. Consequently, at $T>\Tch$, as soon as the contribution to the meson mass coming from the confining interaction exceeds that from the quark kinetic term, the aforementioned additional symmetry should reveal itself as an approximate degeneracy of all states with a given $J>0$. Indeed, one can conclude from Fig.~\ref{fig:grstates} that, in the mesons with $J=1$, chiral spin symmetry is strongly broken at $T=0$ but this breaking is essentially reduced above $\Tch$. At the same time, the emergence of the approximate degeneracy
of all the states with a given spin $J$ is well seen already for
$J\geqslant 2$.

\subsubsection{Wave functions and radii of higher-spin states}

\begin{figure*}[t!]
\centering
\includegraphics[width=0.8\textwidth]{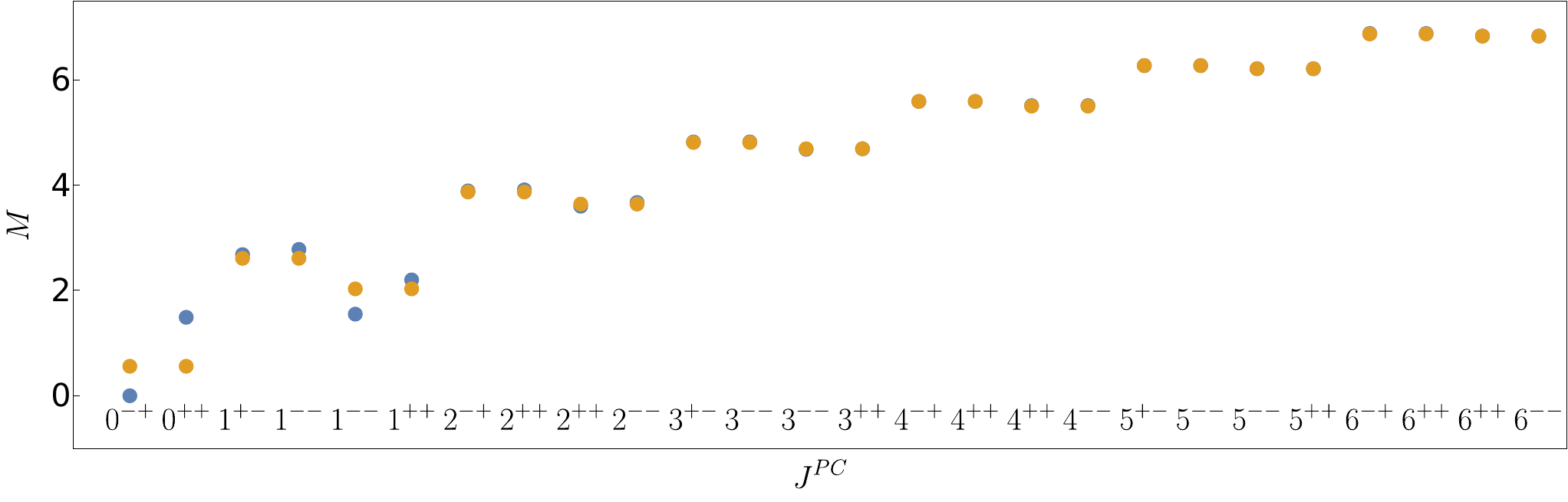}
\caption{Masses (in the units of $\sqrt{\sigma}$) of the  mesons with $n=0$ at $T=0$ (blue) and $T=1.1\Tch$ (yellow).}
\label{fig:grstates}
\scalebox{0.8}{
\begin{tabular}{cc}
\includegraphics[width=0.54\textwidth]{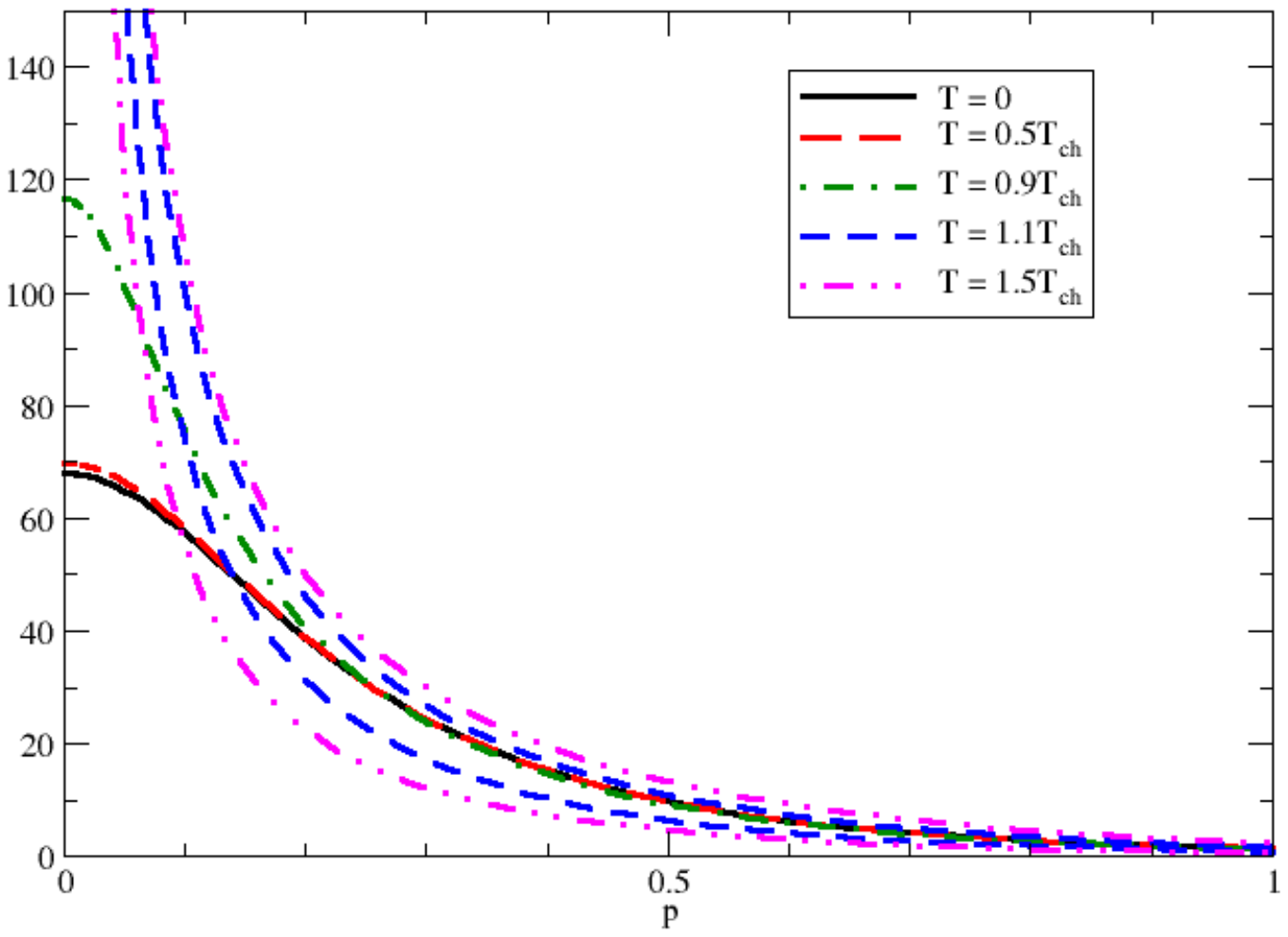}&
\hspace*{-0.08\textwidth}\includegraphics[width=0.54\textwidth]{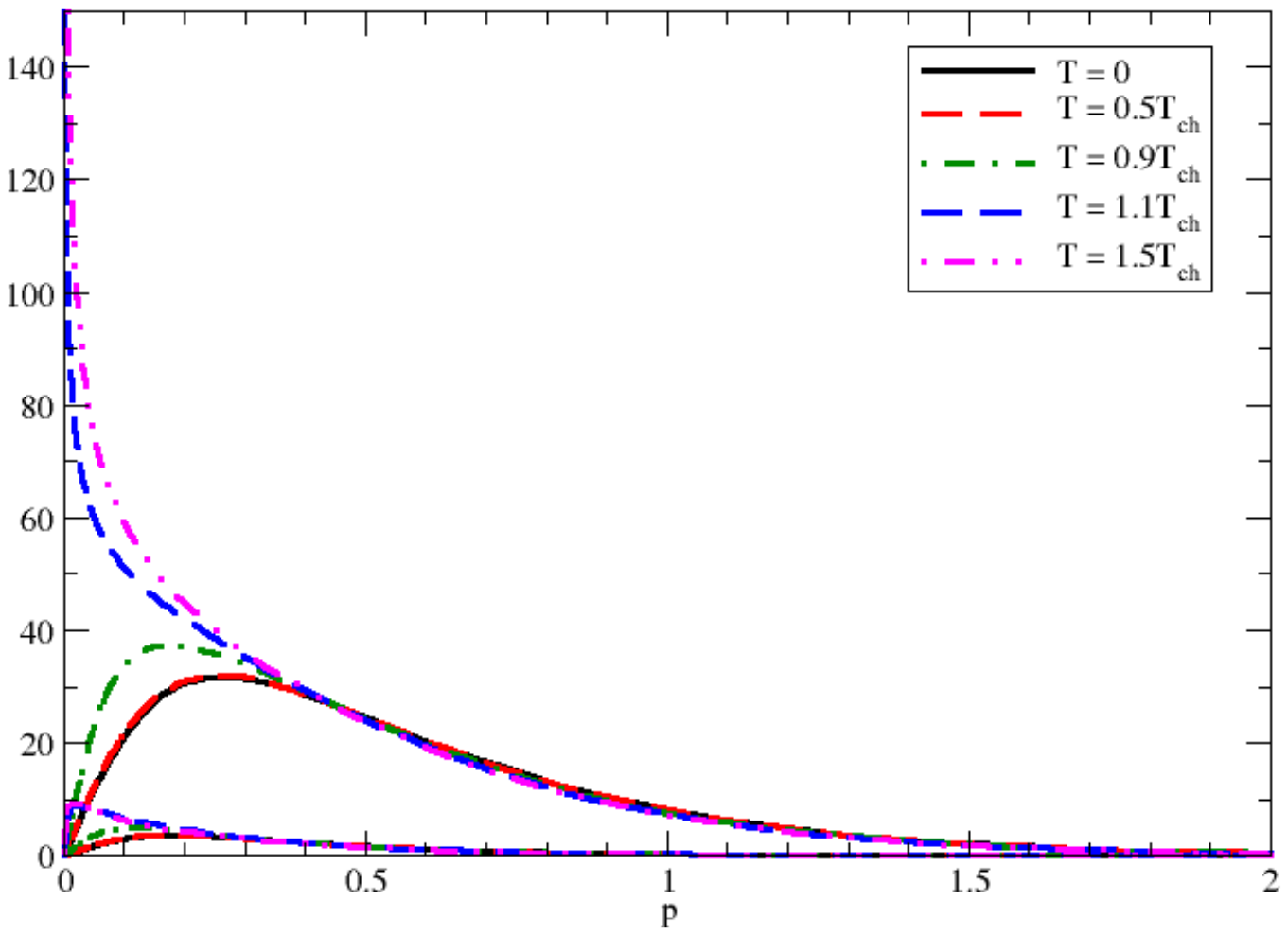}\\[-8mm]
\includegraphics[width=0.54\textwidth]{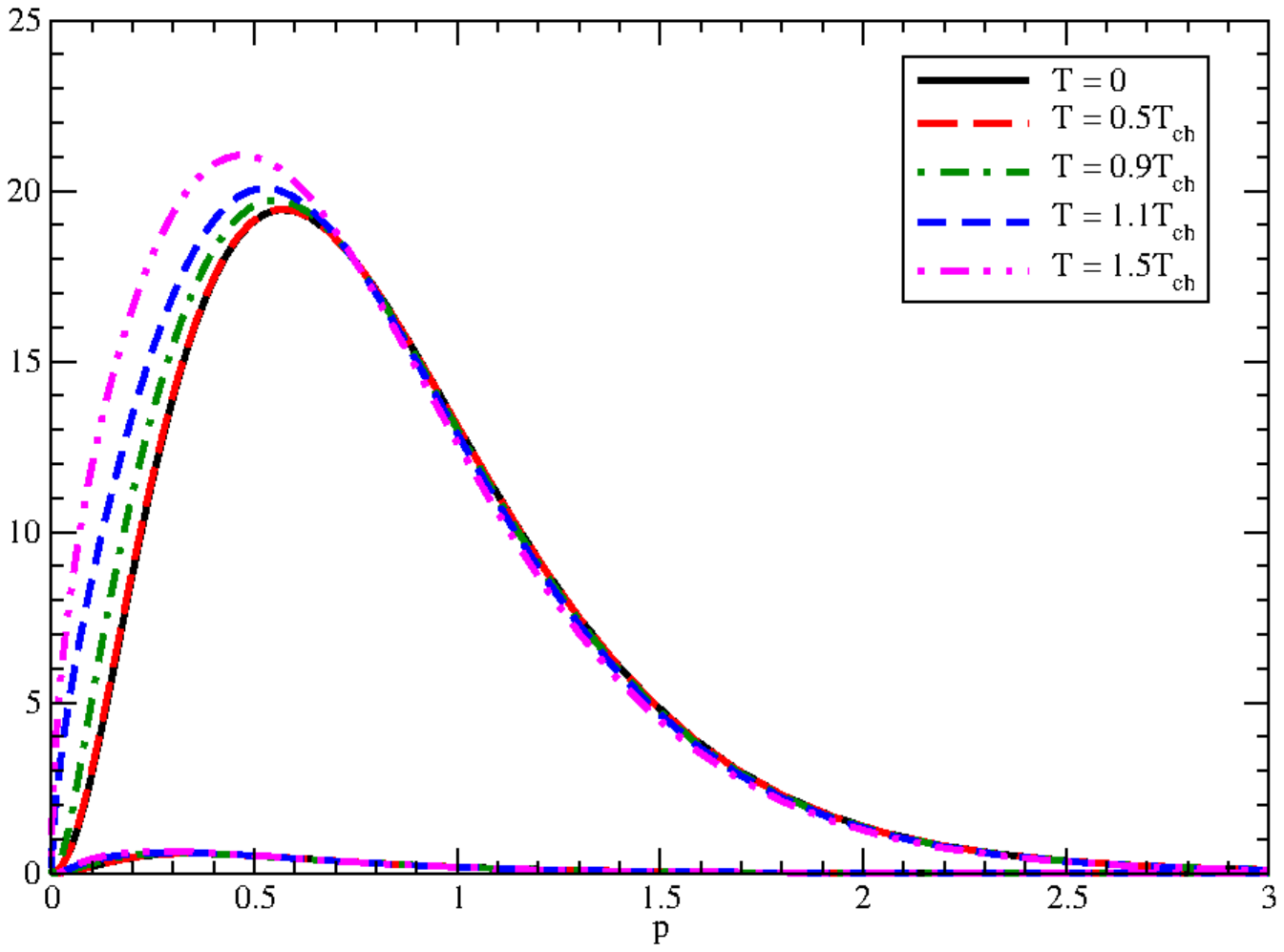}&
\hspace*{-0.02\textwidth}\includegraphics[width=0.54\textwidth]{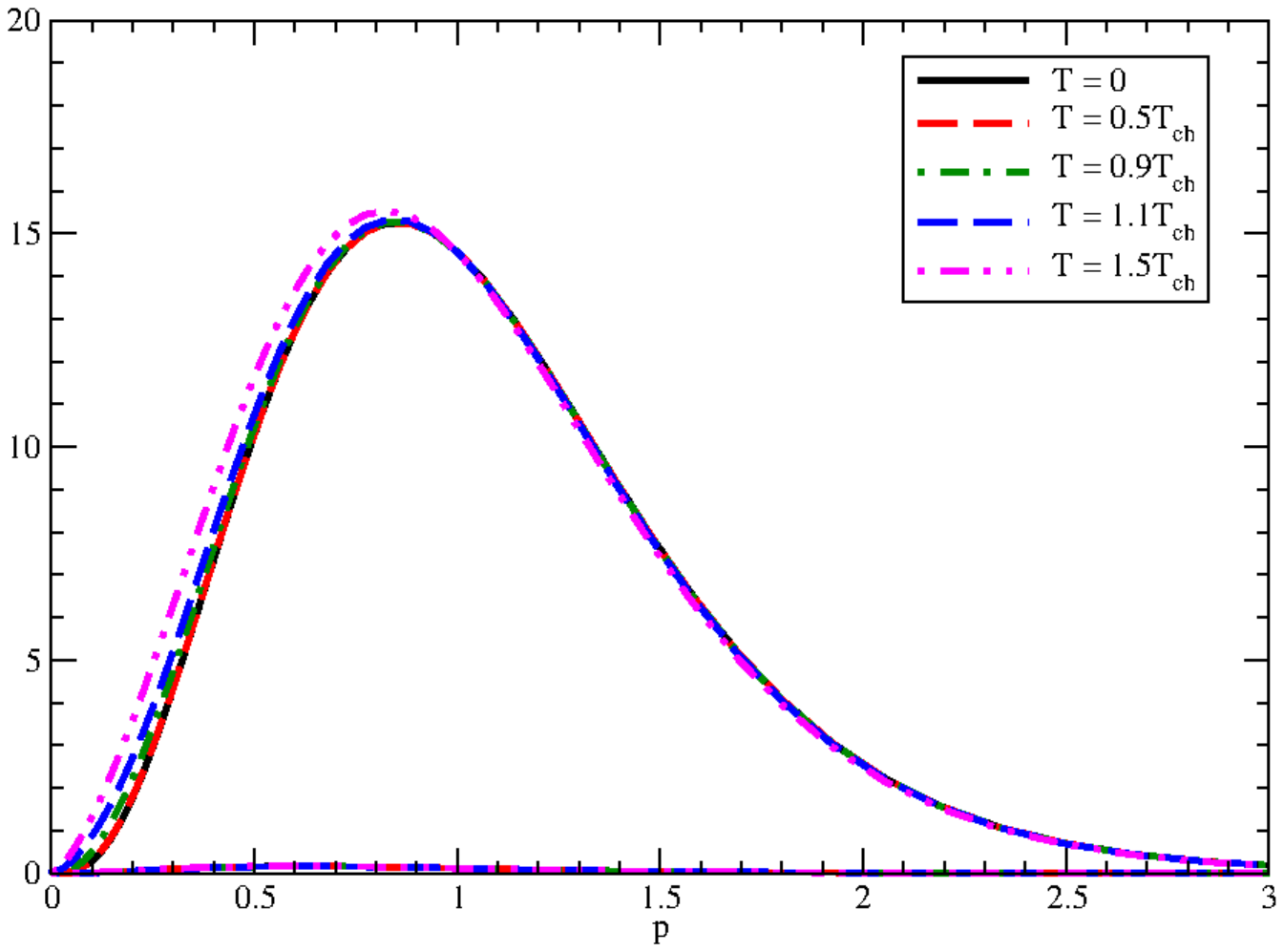}
\end{tabular}}
\caption{The ground state ($n=0$) wave functions $\psi_\pm(p)$ for $J^{PC}=0^{-+}$ (upper left plot), $J^{PC}=1^{+-}$ (upper right plot), $J^{PC}=2^{-+}$ (lower left plot), and $J^{PC}=3^{+-}$ (lower right plot) at different temperatures. For $J^{PC}=0^{-+}$ and $T<\Tch$, the corresponding pseudoscalar meson is a massless Goldstone boson with $\psi_+(p)=\psi_-(p)$. For $J^{PC}=0^{-+}$ and $T>\Tch$ as well as for all other sets of quantum numbers and all temperatures, $\psi_+(p)>\psi_-(p)$, so for each temperature there are two curves in the corresponding plots.}
\label{fig:wfs}
\scalebox{0.8}{\includegraphics[width=0.49\textwidth]{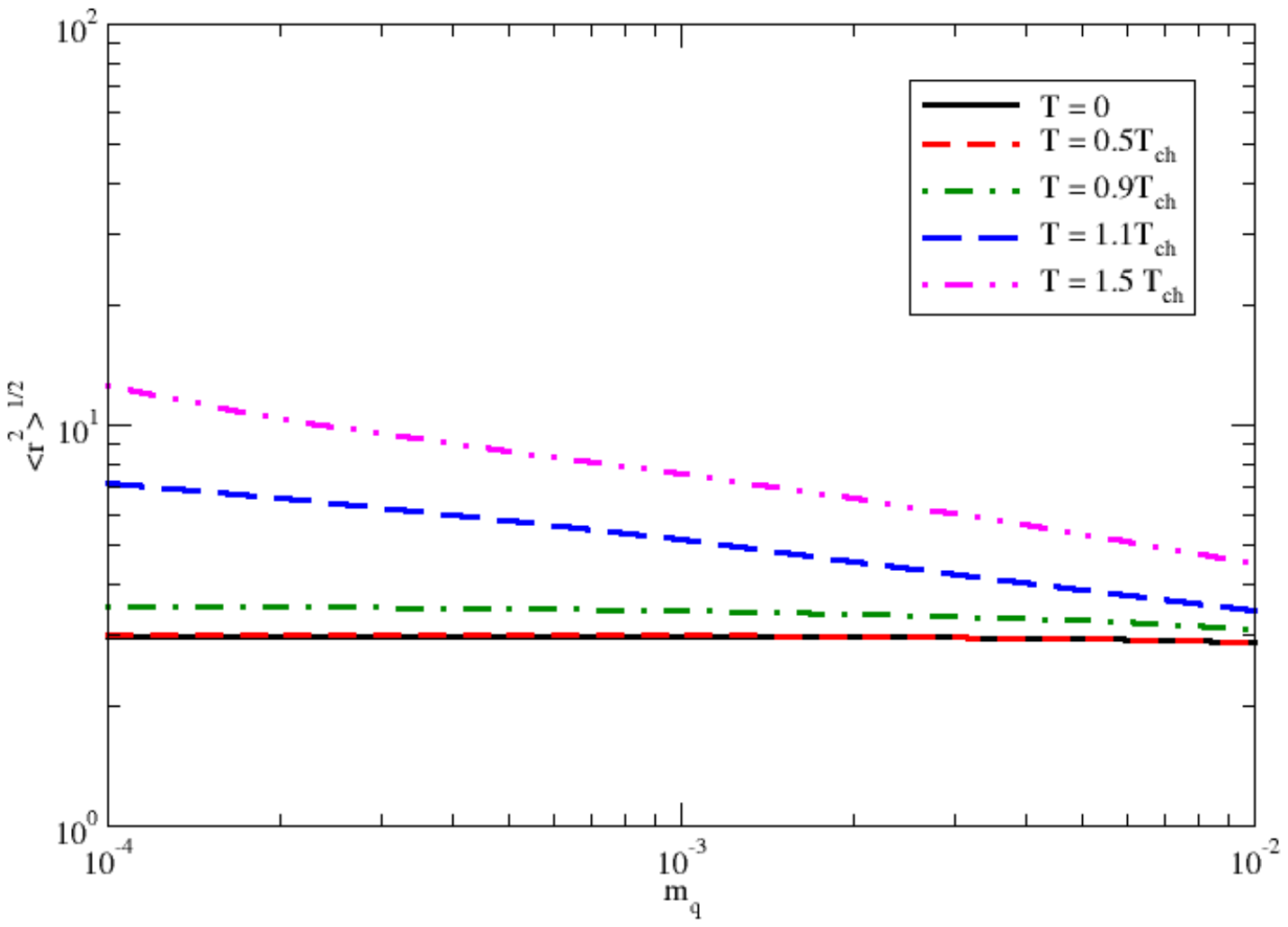}\hspace*{0.03\textwidth}
\includegraphics[width=0.49\textwidth]{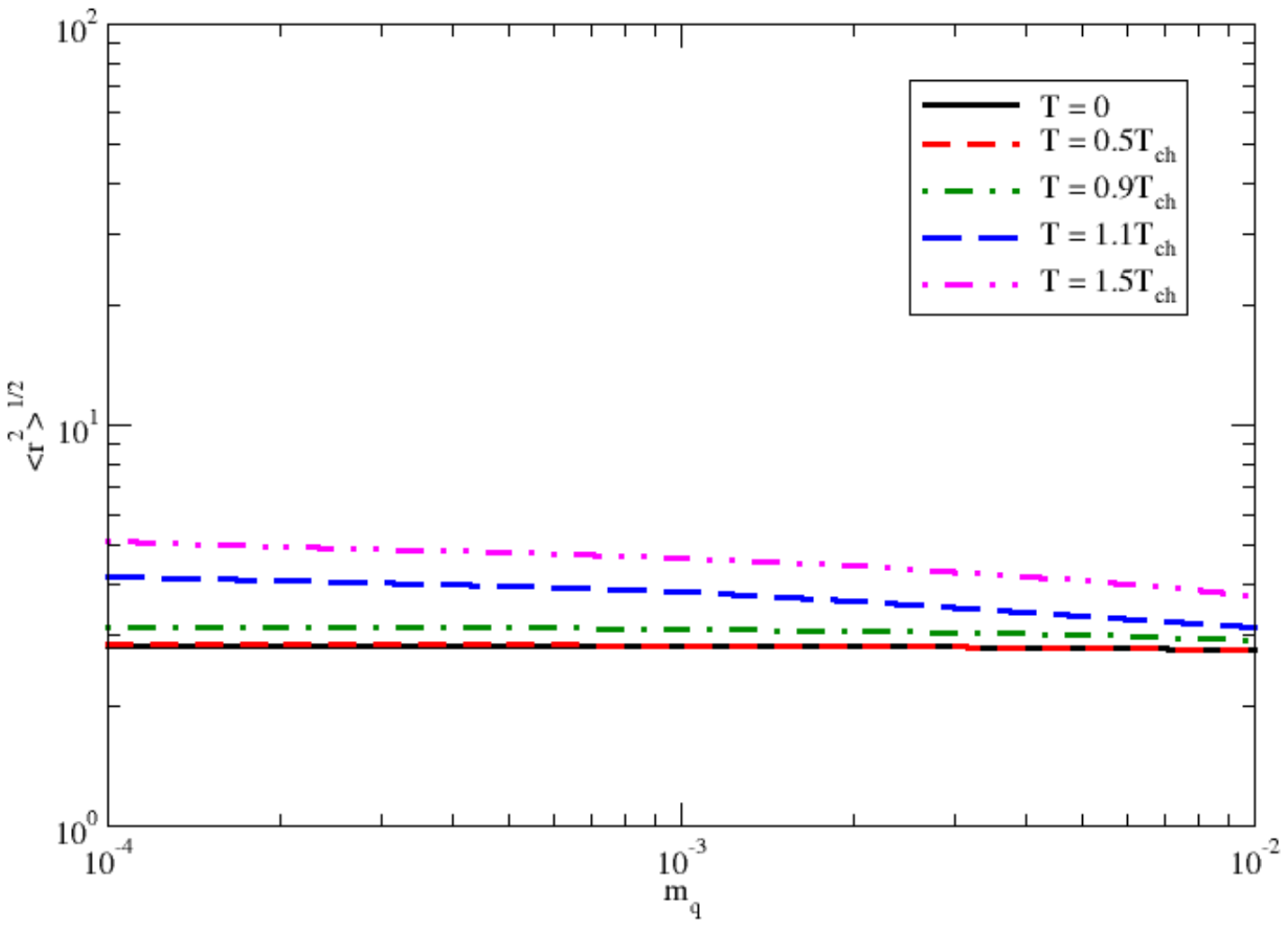}}
\caption{The r.m.s. radius of the ground-state $J^{PC}=1^{+-}$ (first plot) and $J^{PC}=1^{++}$ (second plot) as function of the quark mass at different temperatures.}
\label{fig:radius2}
\end{figure*}

In Fig.~\ref{fig:wfs}, we compare the wave functions of several mesons with different quantum numbers obtained in the chiral limit at different temperatures. In particular, at temperatures $T \lesssim 0.5 \Tch$, the pion wave function $\psi_+^\pi(p;M_\pi=0)=\psi_-^\pi(p,M_\pi=0)$ is approximately the same as at $T=0$. Then, as the temperature  increases towards $\Tch$, the wave function gets enhanced at small momenta and it finally diverges at $p=0$ above the chiral restoration temperature. This observation provides a natural explanation for the divergent pion radius discussed above. Naturally, since chiral symmetry is restored above $\Tch$, the wave functions of the $J^{PC}=0^{-+}$ and $J^{PC}=0^{++}$ states become identical in the chiral limit, so all the conclusions made above for the pion at $T>\Tch$ equally apply to the ``$\sigma$-meson''. A similar, although somewhat weaker, divergence at small momenta is also observed for the wave function of the ``$b_1$-meson'' with
$J^{PC}=1^{+-}$. It can be further deduced from the corresponding plots in the lower row in Fig.~\ref{fig:wfs} that the small-$p$ behaviour of the wave functions for the mesons with higher spins, $J\geqslant 2$, is somewhat tamed with the temperature rise. The difference with the mesons with $J=0,1$ stems from a stronger centrifugal repulsion for higher $J$'s that suppresses the meson wave function at small momenta. Consequently, Pauli blocking of the quark
and antiquark levels at $T>\Tch$ is not as important in the mesons with $J\geqslant 2$ as in their siblings with $J=0,1$. This feature can be further traced by comparing the r.m.s. radii of these mesons. In Table~\ref{tab:rrho}, we provide the results of the calculations of $\braket{r^2}^{1/2}$ for the vector ``$\rho$-meson''. The effect of ``swelling'' at $T>\Tch$ is clearly seen in this case as well, though it is less pronounced than for $J=0$ --- for a realistic
current quark mass, the ``$\rho$-meson'' increases its size at $T>\Tch$ by approximately a factor of 2 (to be compared with $3\div 5$ for the pion and ``$\sigma$-meson'' --- see Tables~\ref{tab:rpi}, \ref{tab:rscalar} and the discussion in the previous subsection). This pattern holds for the other $J=1$ states too as can be seen in Fig.~\ref{fig:radius2}:
although the radii of the $J^{PC}=1^{+-}$ and $J^{PC}=1^{++}$ states also demonstrate an unbound rise in the approach to the chiral limit, it is not as strong as in the case of the mesons with $J=0$ --- \emph{cf.} Fig.~\ref{fig:radius}. This rise is further tamed for the states with higher spins $J$.

We, therefore, conclude that a dilute gas of hadrons, observed at temperatures below $\Tch$, turns to a dense system of swelled, strongly overlapping meson-like states with light quarks (predominantly with $J=0,1$) above the chiral restoration temperature, and the corresponding phase of the theory deserves its name ``stringy fluid''.

\section{Conclusions}

In this paper, we employed a manifestly confining and chirally
symmetric model for QCD to study some properties of the hot QCD
matter above the chiral restoration temperature. This model is only an approximation to QCD since all interquark interactions, except for the linear confining potential, are disregarded.
Nevertheless, it accommodates the principal aspects of QCD at low temperatures such as confinement and spontaneous breaking of chiral symmetry in the vacuum. It also shares chiral spin symmetry of confinement with real QCD. Thus we dare expect this model to provide a valuable microscopic insight into hot QCD as well.

We observed a chiral symmetry restoration phase transition at $\Tch \simeq 90$ MeV that should be confronted with the critical temperature around 130 MeV obtained on the lattice in the chiral limit. The physical reason for chiral restoration in the
confining regime is Pauli blocking of the fermionic levels, necessary for the formation of the quark condensate, by the thermal excitation of quarks and antiquarks. Since the model remains confining above $\Tch$, its spectrum still consists of hadron-like states rather than free quarks. We solved the bound-state equation for quark-antiquark mesons at $T>\Tch$ to find their masses and wave functions near the chiral limit of $\mq\to 0$. A crucial property of these meson-like states with light quarks is their size (estimated from the r.m.s. radius) that, for a realistic current quark mass, may exceed several times the corresponding size at zero temperature. Furthermore, their radii diverge in the chiral limit thus driving the corresponding bound states infinitely large. The physics behind this swelling is, again, Pauli blocking of the fermionic levels with small momenta by thermal excitations. Importantly, even though the size of these hadron-like objects becomes (infinitely) large, they are still quark-antiquark bound states with well defined masses and normalisable wave functions. The most pronounced effect is observed for lower spins $J$ since Pauli blocking stronger affects physical systems most sensitive to small momenta. Mesons consisting of heavy quarks are expected to preserve their size above $\Tch$ since Pauli blocking due to the thermal excitations is not operative in them.

We stress that deconfined quarks and antiquarks can not exist in the employed model since it remains confining at all temperatures. Therefore, a dilute hadron gas below $\Tch$ converts into a system of very large overlapping meson-like states (``strings'') above the chiral restoration temperature. The observed radical increase in size of the meson-like states at $T>\Tch$ naturally explains a crucial experimental property of the hot QCD matter --- namely, its collectivity and a very small mean free path of the effective constituents.

\section*{Acknowledgments}

Work of A.N. was supported by Deutsche Forschungsgemeinschaft (Project No. 525056915). He also acknowledges support from the CAS President’s International Fellowship Initiative (Grant No. 2024PVA0004) during his stay in China.


\providecommand{\href}[2]{#2}\begingroup\raggedright\endgroup



\clearpage

\begin{table*}
\centering
\caption{Masses (in the units of $\sqrt{\sigma}$) of isovector mesons at $T=0$.}
\begin{tabular}{|c|cccccccc|}
\hline\hline
Chiral\hspace*{5mm}&\raisebox{-1.5ex}{$J^{PC}$}&
\multicolumn{7}{c|}{Radial excitation $n$}\\[-1.5ex]
multiplet&&0&1&2&3&
4&5&6\\
\hline
$(1/2,1/2)_a$&$0^{-+}$&0.00&2.93&4.35&5.49&6.46&7.31&8.09\\
$(1/2,1/2)_b$&$0^{++}$&1.49&3.38&4.72&5.80&6.74&7.57&8.33\\
\hline
$(1/2,1/2)_a$&$1^{+-}$&2.68&4.03&5.16&6.14&7.01&7.80&8.53\\
$(1/2,1/2)_b$&$1^{--}$&2.78&4.18&5.32&6.30&7.17&7.96&8.68\\
$(0,1)\oplus(1,0)$&$1^{--}$&1.55&3.28&4.56&5.64&6.57&7.40&8.16\\
$(0,1)\oplus(1,0)$&$1^{++}$&2.20&3.73&4.95&5.98&6.88&7.69&8.43\\
\hline
$(1/2,1/2)_a$&$2^{-+}$&3.89&4.98&5.94&6.80&7.59&8.31&8.99\\
$(1/2,1/2)_b$&$2^{++}$&3.91&5.02&6.00&6.88&7.67&8.40&9.09\\
$(0,1)\oplus(1,0)$&$2^{++}$&3.60&4.67&5.64&6.51&7.32&8.06&8.75\\
$(0,1)\oplus(1,0)$&$2^{--}$&3.67&4.80&5.80&6.68&7.49&8.23&8.91\\
\hline
$(1/2,1/2)_a$&$3^{+-}$&4.82&5.77&6.62&7.41&8.13&8.81&9.45\\
$(1/2,1/2)_b$&$3^{--}$&4.82&5.78&6.65&7.44&8.17&8.86&9.50\\
$(0,1)\oplus(1,0)$&$3^{--}$&4.68&5.63&6.48&7.26&7.99&8.67&9.30\\
$(0,1)\oplus(1,0)$&$3^{++}$&4.69&5.66&6.53&7.32&8.06&8.74&9.39\\
\hline
$(1/2,1/2)_a$&$4^{-+}$&5.59&6.45&7.23&7.96&8.64&9.28&9.89\\
$(1/2,1/2)_b$&$4^{++}$&5.59&6.45&7.24&7.97&8.66&9.30&9.91\\
$(0,1)\oplus(1,0)$&$4^{++}$&5.51&6.36&7.15&7.88&8.56&9.19&9.80\\
$(0,1)\oplus(1,0)$&$4^{--}$&5.51&6.37&7.16&7.90&8.58&9.23&9.84\\
\hline
$(1/2,1/2)_a$&$5^{+-}$&6.27&7.05&7.78&8.47&9.11&9.72&10.3\\
$(1/2,1/2)_b$&$5^{--}$&6.27&7.06&7.79&8.47&9.12&9.73&10.3\\
$(0,1)\oplus(1,0)$&$5^{--}$&6.21&7.00&7.73&8.41&9.06&9.67&10.2\\
$(0,1)\oplus(1,0)$&$5^{++}$&6.21&7.00&7.73&8.42&9.07&9.68&10.3\\
\hline
$(1/2,1/2)_a$&$6^{-+}$&6.88&7.61&8.29&8.94&9.55&10.1&10.7\\
$(1/2,1/2)_b$&$6^{++}$&6.88&7.61&8.29&8.94&9.56&10.1&10.7\\
$(0,1)\oplus(1,0)$&$6^{++}$&6.83&7.57&8.25&8.90&9.51&10.1&10.7\\
$(0,1)\oplus(1,0)$&$6^{--}$&6.83&7.57&8.26&8.90&9.52&10.1&10.7\\
\hline\hline
\end{tabular}
\label{tab:spect0}
\end{table*}

\begin{table*}
\centering
\caption{Masses (in the units of $\sqrt{\sigma}$) of isovector mesons at $T=0.5\Tch$.}
\begin{tabular}{|c|cccccccc|}
\hline\hline
Chiral\hspace*{5mm}&\raisebox{-1.5ex}{$J^{PC}$}&
\multicolumn{7}{c|}{Radial excitation $n$}\\[-1.5ex]
multiplet&&0&1&2&3&
4&5&6\\
\hline
$(1/2,1/2)_a$&$0^{-+}$&0.00&2.93&4.35&5.48&6.45&7.31&8.08\\
$(1/2,1/2)_b$&$0^{++}$&1.47&3.37&4.71&5.79&6.73&7.56&8.32\\
\hline
$(1/2,1/2)_a$&$1^{+-}$&2.68&4.02&5.16&6.14&7.01&7.8&8.52\\
$(1/2,1/2)_b$&$1^{--}$&2.77&4.17&5.32&6.3&7.16&7.95&8.67\\
$(0,1)\oplus(1,0)$&$1^{--}$&1.57&3.28&4.56&5.63&6.57&7.4&8.16\\
$(0,1)\oplus(1,0)$&$1^{++}$&2.20&3.72&4.94&5.97&6.87&7.68&8.42\\
\hline
$(1/2,1/2)_a$&$2^{-+}$&3.89&4.97&5.93&6.80&7.58&8.31&8.99\\
$(1/2,1/2)_b$&$2^{++}$&3.91&5.02&6.00&6.87&7.67&8.40&9.08\\
$(0,1)\oplus(1,0)$&$2^{++}$&3.60&4.68&5.64&6.51&7.31&8.05&8.74\\
$(0,1)\oplus(1,0)$&$2^{--}$&3.67&4.80&5.79&6.68&7.48&8.22&8.91\\
\hline
$(1/2,1/2)_a$&$3^{+-}$&4.82&5.77&6.62&7.41&8.13&8.81&9.45\\
$(1/2,1/2)_b$&$3^{--}$&4.82&5.78&6.65&7.44&8.17&8.86&9.50\\
$(0,1)\oplus(1,0)$&$3^{--}$&4.68&5.63&6.48&7.26&7.99&8.67&9.30\\
$(0,1)\oplus(1,0)$&$3^{++}$&4.69&5.66&6.53&7.32&8.06&8.74&9.39\\
\hline
$(1/2,1/2)_a$&$4^{-+}$&5.59&6.45&7.23&7.96&8.64&9.28&9.89\\
$(1/2,1/2)_b$&$4^{++}$&5.59&6.45&7.24&7.97&8.66&9.30&9.91\\
$(0,1)\oplus(1,0)$&$4^{++}$&5.50&6.36&7.15&7.88&8.55&9.19&9.80\\
$(0,1)\oplus(1,0)$&$4^{--}$&5.51&6.37&7.16&7.89&8.58&9.23&9.83\\
\hline
$(1/2,1/2)_a$&$5^{+-}$&6.27&7.06&7.78&8.47&9.11&9.72&10.3\\
$(1/2,1/2)_b$&$5^{--}$&6.27&7.06&7.79&8.47&9.06&9.73&10.3\\
$(0,1)\oplus(1,0)$&$5^{--}$&6.21&7.00&7.73&8.41&9.06&9.67&10.2\\
$(0,1)\oplus(1,0)$&$5^{++}$&6.21&7.00&7.73&8.42&9.07&9.68&10.3\\
\hline
$(1/2,1/2)_a$&$6^{-+}$&6.87&7.60&8.29&8.93&9.55&10.1&10.7\\
$(1/2,1/2)_b$&$6^{++}$&6.87&7.60&8.29&8.94&9.55&10.1&10.7\\
$(0,1)\oplus(1,0)$&$6^{++}$&6.83&7.56&8.25&8.90&9.51&10.1&10.7\\
$(0,1)\oplus(1,0)$&$6^{--}$&6.83&7.56&8.25&8.90&9.51&10.1&10.7\\
\hline\hline
\end{tabular}
\label{tab:spect05tch}
\end{table*}

\begin{table*}
\centering
\caption{Masses (in the units of $\sqrt{\sigma}$) of isovector mesons at $T=0.9\Tch$.}
\begin{tabular}{|c|cccccccc|}
\hline\hline
Chiral\hspace*{5mm}&\raisebox{-1.5ex}{$J^{PC}$}&
\multicolumn{7}{c|}{Radial excitation $n$}\\[-1.5ex]
multiplet&&0&1&2&3&
4&5&6\\
\hline
$(1/2,1/2)_a$&$0^{-+}$&0.00&2.88&4.27&5.38&6.33&7.17&7.93\\
$(1/2,1/2)_b$&$0^{++}$&1.15&3.18&4.53&5.62&6.56&7.39&8.15\\
\hline
$(1/2,1/2)_a$&$1^{+-}$&2.64&3.96&5.08&6.05&6.91&7.69&8.40\\
$(1/2,1/2)_b$&$1^{--}$&2.69&4.05&5.18&6.16&7.02&7.81&8.53\\
$(0,1)\oplus(1,0)$&$1^{--}$&1.78&3.32&4.54&5.57&6.47&7.29&8.03\\
$(0,1)\oplus(1,0)$&$1^{++}$&2.10&3.62&4.83&5.85&6.74&7.55&8.29\\
\hline
$(1/2,1/2)_a$&$2^{-+}$&3.88&4.95&5.9&6.75&7.53&8.25&8.92\\
$(1/2,1/2)_b$&$2^{++}$&3.89&4.98&5.94&6.80&7.59&8.31&8.99\\
$(0,1)\oplus(1,0)$&$2^{++}$&3.63&4.71&5.66&6.51&7.30&8.02&8.70\\
$(0,1)\oplus(1,0)$&$2^{--}$&3.66&4.77&5.74&6.62&7.41&8.14&8.83\\
\hline
$(1/2,1/2)_a$&$3^{+-}$&4.82&5.76&6.61&7.39&8.11&8.79&9.42\\
$(1/2,1/2)_b$&$3^{--}$&4.82&5.77&6.62&7.41&8.13&8.81&9.45\\
$(0,1)\oplus(1,0)$&$3^{--}$&4.69&5.64&6.50&7.28&8.00&8.67&9.30\\
$(0,1)\oplus(1,0)$&$3^{++}$&4.69&5.65&6.52&7.30&8.03&8.71&9.35\\
\hline
$(1/2,1/2)_a$&$4^{-+}$&5.59&6.44&7.23&7.95&8.63&9.27&9.88\\
$(1/2,1/2)_b$&$4^{++}$&5.59&6.45&7.23&7.96&8.64&9.208&9.89\\
$(0,1)\oplus(1,0)$&$4^{++}$&5.50&6.37&7.15&7.88&8.56&9.2&9.81\\
$(0,1)\oplus(1,0)$&$4^{--}$&5.51&6.37&7.16&7.89&8.57&9.21&9.82\\
\hline
$(1/2,1/2)_a$&$5^{+-}$&6.27&7.06&7.78&8.47&9.11&9.72&10.3\\
$(1/2,1/2)_b$&$5^{--}$&6.27&7.06&7.79&8.47&9.11&9.72&10.3\\
$(0,1)\oplus(1,0)$&$5^{--}$&6.21&7.00&7.73&8.42&9.06&9.67&10.3\\
$(0,1)\oplus(1,0)$&$5^{++}$&6.21&7.00&7.73&8.42&9.07&9.68&10.3\\
\hline
$(1/2,1/2)_a$&$6^{-+}$&6.87&7.60&8.29&8.93&9.55&10.1&10.7\\
$(1/2,1/2)_b$&$6^{++}$&6.87&7.60&8.29&8.93&9.55&10.1&10.7\\
$(0,1)\oplus(1,0)$&$6^{++}$&6.83&7.56&8.25&8.90&9.51&10.1&10.7\\
$(0,1)\oplus(1,0)$&$6^{--}$&6.83&7.56&8.25&8.90&9.51&10.1&10.7\\
\hline\hline
\end{tabular}
\label{tab:spect09tch}
\end{table*}

\begin{table*}
\centering
\caption{Masses (in the units of $\sqrt{\sigma}$) of isovector mesons at $T=1.1\Tch$.}
\begin{tabular}{|c|cccccccc|}
\hline\hline
Chiral\hspace*{5mm}&\raisebox{-1.5ex}{$J^{PC}$}&
\multicolumn{7}{c|}{Radial excitation $n$}\\[-1.5ex]
multiplet&&0&1&2&3&
4&5&6\\
\hline
$(1/2,1/2)_a$&$0^{-+}$&0.56&2.83&4.09&5.04&5.81&6.51&7.19\\
$(1/2,1/2)_b$&$0^{++}$&0.56&2.83&4.09&5.04&5.81&6.51&7.19\\
\hline
$(1/2,1/2)_a$&$1^{+-}$&2.61&3.91&4.99&5.92&6.73&7.46&8.12\\
$(1/2,1/2)_b$&$1^{--}$&2.61&3.91&4.99&5.92&6.73&7.46&8.12\\
$(0,1)\oplus(1,0)$&$1^{--}$&2.03&3.49&4.64&5.60&6.41&7.14&7.79\\
$(0,1)\oplus(1,0)$&$1^{++}$&2.03&3.49&4.64&5.60&6.41&7.14&7.79\\
\hline
$(1/2,1/2)_a$&$2^{-+}$&3.87&4.93&5.86&6.70&7.47&8.17&8.82\\
$(1/2,1/2)_b$&$2^{++}$&3.87&4.93&5.86&6.70&7.47&8.17&8.82\\
$(0,1)\oplus(1,0)$&$2^{++}$&3.64&4.74&5.69&6.55&7.32&8.03&8.69\\
$(0,1)\oplus(1,0)$&$2^{--}$&3.64&4.74&5.69&6.55&7.32&8.03&8.69\\
\hline
$(1/2,1/2)_a$&$3^{+-}$&4.81&5.75&6.60&7.37&8.09&8.75&9.38\\
$(1/2,1/2)_b$&$3^{--}$&4.81&5.75&6.60&7.37&8.09&8.75&9.38\\
$(0,1)\oplus(1,0)$&$3^{--}$&4.69&5.64&6.50&7.28&8.00&8.67&9.30\\
$(0,1)\oplus(1,0)$&$3^{++}$&4.69&5.64&6.50&7.28&8.00&8.67&9.30\\
\hline
$(1/2,1/2)_a$&$4^{-+}$&5.59&6.44&7.22&7.95&8.62&9.26&9.86\\
$(1/2,1/2)_b$&$4^{++}$&5.59&6.44&7.22&7.95&8.62&9.26&9.86\\
$(0,1)\oplus(1,0)$&$4^{++}$&5.50&6.37&7.15&7.88&8.56&9.20&9.80\\
$(0,1)\oplus(1,0)$&$4^{--}$&5.50&6.37&7.15&7.88&8.56&9.20&9.80\\
\hline
$(1/2,1/2)_a$&$5^{+-}$&6.27&7.05&7.78&8.46&9.10&9.71&10.3\\
$(1/2,1/2)_b$&$5^{--}$&6.27&7.05&7.78&8.46&9.10&9.71&10.3\\
$(0,1)\oplus(1,0)$&$5^{--}$&6.21&7.00&7.73&8.42&9.06&9.67&10.2\\
$(0,1)\oplus(1,0)$&$5^{++}$&6.21&7.00&7.73&8.42&9.06&9.67&10.2\\
\hline
$(1/2,1/2)_a$&$6^{-+}$&6.87&7.60&8.29&8.93&9.54&10.1&10.7\\
$(1/2,1/2)_b$&$6^{++}$&6.87&7.60&8.29&8.93&9.54&10.1&10.7\\
$(0,1)\oplus(1,0)$&$6^{++}$&6.83&7.56&8.25&8.90&9.51&10.1&10.7\\
$(0,1)\oplus(1,0)$&$6^{--}$&6.83&7.56&8.25&8.90&9.51&10.1&10.7\\
\hline\hline
\end{tabular}
\label{tab:spect11tch}
\end{table*}

\begin{table*}
\centering
\caption{Masses (in the units of $\sqrt{\sigma}$) of isovector mesons at $T=1.5\Tch$.}
\begin{tabular}{|c|cccccccc|}
\hline\hline
Chiral\hspace*{5mm}&\raisebox{-1.5ex}{$J^{PC}$}&
\multicolumn{7}{c|}{Radial excitation $n$}\\[-1.5ex]
multiplet&&0&1&2&3&
4&5&6\\
\hline
$(1/2,1/2)_a$&$0^{-+}$&1.08&2.85&3.86&4.61&5.34&6.08&6.79\\
$(1/2,1/2)_b$&$0^{++}$&1.08&2.85&3.86&4.61&5.34&6.08&6.79\\
\hline
$(1/2,1/2)_a$&$1^{+-}$&2.62&3.88&4.90&5.74&6.45&7.06&7.63\\
$(1/2,1/2)_b$&$1^{--}$&2.62&3.88&4.90&5.74&6.45&7.06&7.63\\
$(0,1)\oplus(1,0)$&$1^{--}$&2.09&3.47&4.52&5.34&6.03&6.69&7.33\\
$(0,1)\oplus(1,0)$&$1^{++}$&2.09&3.47&4.52&5.34&6.03&6.69&7.33\\
\hline
$(1/2,1/2)_a$&$2^{-+}$&3.85&4.89&5.80&6.61&7.35&8.02&8.63\\
$(1/2,1/2)_b$&$2^{++}$&3.85&4.89&5.80&6.61&7.35&8.02&8.63\\
$(0,1)\oplus(1,0)$&$2^{++}$&3.62&4.69&5.62&6.45&7.19&7.85&8.46\\
$(0,1)\oplus(1,0)$&$2^{--}$&3.62&4.69&5.62&6.45&7.19&7.85&8.46\\
\hline
$(1/2,1/2)_a$&$3^{+-}$&4.80&5.73&6.56&7.32&8.02&8.67&9.28\\
$(1/2,1/2)_b$&$3^{--}$&4.80&5.73&6.56&7.32&8.02&8.67&9.28\\
$(0,1)\oplus(1,0)$&$3^{--}$&4.68&5.61&6.46&7.22&7.93&8.58&9.20\\
$(0,1)\oplus(1,0)$&$3^{++}$&4.68&5.61&6.46&7.22&7.93&8.58&9.20\\
\hline
$(1/2,1/2)_a$&$4^{-+}$&5.58&6.43&7.20&7.92&8.58&9.21&9.80\\
$(1/2,1/2)_b$&$4^{++}$&5.58&6.43&7.20&7.92&8.58&9.21&9.80\\
$(0,1)\oplus(1,0)$&$4^{++}$&5.50&6.35&7.13&7.85&8.52&9.15&9.74\\
$(0,1)\oplus(1,0)$&$4^{--}$&5.50&6.35&7.13&7.85&8.52&9.15&9.74\\
\hline
$(1/2,1/2)_a$&$5^{+-}$&6.27&7.05&7.77&8.45&9.08&9.69&10.3\\
$(1/2,1/2)_b$&$5^{--}$&6.27&7.05&7.77&8.45&9.08&9.69&10.3\\
$(0,1)\oplus(1,0)$&$5^{--}$&6.21&6.99&7.72&8.40&9.04&9.64&10.2\\
$(0,1)\oplus(1,0)$&$5^{++}$&6.21&6.99&7.72&8.40&9.04&9.64&10.2\\
\hline
$(1/2,1/2)_a$&$6^{-+}$&6.87&7.60&8.28&8.92&9.53&10.1&10.7\\
$(1/2,1/2)_b$&$6^{++}$&6.87&7.60&8.28&8.92&9.53&10.1&10.7\\
$(0,1)\oplus(1,0)$&$6^{++}$&6.83&7.56&8.25&8.89&9.50&10.1&10.6\\
$(0,1)\oplus(1,0)$&$6^{--}$&6.83&7.56&8.25&8.89&9.50&10.1&10.6\\
\hline\hline
\end{tabular}
\label{tab:spect15tch}
\end{table*}

\begin{table*}[t!]
\caption{The r.m.s. radius $\left<r^2\right>^{1/2}$ of the ``$\rho$-meson'' in the units of $\sigma^{-1/2}$.}
\label{tab:rrho}
\centering
\begin{tabular}{|c|cccccccc|}
\hline\hline
$T\backslash \mq$&0&$10^{-4}$&$2\cdot10^{-4}$&$5\cdot10^{-4}$&$10^{-3}$&$2\cdot10^{-3}$&$5\cdot10^{-3}$&$10^{-2}$\\\hline
0&4.32&4.31&4.31&4.30&4.29&4.26&4.19&4.07\\
$0.5\Tch$&4.35&4.35&4.34&4.33&4.32&4.29&4.21&4.09\\
$0.9\Tch$&5.02&5.01&5.00&4.97&4.92&4.83&4.63&3.40\\
$1.1\Tch$&$\infty$&7.99&7.36&6.65&6.21&5.84&5.32&4.85\\
$1.5\Tch$&$\infty$&14.6&12.1&10.1&8.95&7.96&6.86&6.08\\
\hline\hline
\end{tabular}
\end{table*}

\end{document}